%% file: main.tex
\documentclass[sigconf, nonacm]{acmart}

\newcommand\vldbyear{2026}
\newcommand\vldbworkshop{Applied AI for Database Systems and
Applications (AIDB 2026)}
\newcommand\vldbauthors{\authors}
\newcommand\vldbtitle{\shorttitle} 
\newcommand\vldbavailabilityurl{https://github.com/mrateike/pvldb-submission}
\newcommand\vldbpagestyle{plain}

\input{format/packages}
\input{format/commands_general}

\input{format/commands_paper}

\usepackage{balance}

\begin{document}

\title{Stage‑Level Executor Allocation in \Apache\ \\ with Cost–Performance Trade‑offs}

\author{Miriam Rateike}
\email{miriam.rateike@ibm.com}
\affiliation{%
  \institution{IBM, University of Tübingen}
  \city{Nairobi}
  \country{Kenya}
}

\author{Isaac Waweru Wambugu}
\email{isaacw@ke.ibm.com}
\affiliation{%
  \institution{IBM}
  \city{Nairobi}
  \country{Kenya}
}

\author{Celia Cintas}
\email{celia.cintas@ibm.com}
\affiliation{%
  \institution{IBM}
  \city{Nairobi}
  \country{Kenya}
}

\author{Michael Kaufmann}
\email{kau@zurich.ibm.com}
\affiliation{%
  \institution{IBM}
  \city{Zurich}
  \country{Switzerland}
}

\author{Ioana Giurgiu}
\email{igi@zurich.ibm.com}
\affiliation{%
  \institution{IBM}
  \city{Zurich}
  \country{Switzerland}
}

\author{Skyler Speakman}
\email{skyler@ke.ibm.com}
\affiliation{%
  \institution{IBM}
  \city{Nairobi}
  \country{Kenya}
}

\renewcommand{\shortauthors}{Rateike et al.}

\input{sections/abstract}

\maketitle

\title{Stage‑Level Executor Allocation in \Apache\ with Cost–Performance Trade‑offs}

\pagestyle{\vldbpagestyle}
\begingroup\small\noindent\raggedright\textbf{VLDB Workshop Reference Format:}\\
\vldbauthors. \vldbtitle. VLDB \vldbyear\ Workshop: \vldbworkshop.\\ 
\endgroup
\begingroup
\renewcommand\thefootnote{}\footnote{\noindent
This work is licensed under the Creative Commons BY-NC-ND 4.0 International License. Visit \url{https://creativecommons.org/licenses/by-nc-nd/4.0/} to view a copy of this license. For any use beyond those covered by this license, obtain permission by emailing \href{mailto:info@vldb.org}{info@vldb.org}. Copyright is held by the owner/author(s). Publication rights licensed to the VLDB Endowment. \\
\raggedright Proceedings of the VLDB Endowment.
ISSN 2150-8097. \\
}\addtocounter{footnote}{-1}\endgroup

\ifdefempty{\vldbavailabilityurl}{}{
\vspace{.3cm}
\begingroup\small\noindent\raggedright\textbf{VLDB Workshop Artifact Availability:}\\
The source code, data, and/or other artifacts have been made available at \url{\vldbavailabilityurl}.
\endgroup
}

\input{sections/introduction}

\input{sections/background}
\input{sections/design-implementation}

\input{sections/results}
\input{sections/discussion}
\input{sections/summary}

\bibliographystyle{ACM-Reference-Format}
\bibliography{references}

\input{sections/appendix}

\input{sections/appendix-a-model-details}

\input{sections/appendix-b-baselines}

\input{sections/appendix-c-extended-results}

\input{sections/appendix-d-hardware-setup}

\end{document}

%% file: format/packages.tex
\usepackage{algorithm}         % Algorithm environment
\usepackage{algpseudocode}     % Algorithmic pseudocode

\usepackage{multirow}

\usepackage{subcaption} % for subfigure environments and subcaptions

%% file: format/commands_general.tex
\everypar{\looseness=-1}
\usepackage{svg}

\theoremstyle{definition}
\newtheorem{definition}{Definition}

\definecolor{bleudefrance}{rgb}{0.19, 0.55, 0.91}
\definecolor{asparagus}{rgb}{0.53, 0.66, 0.42}
\definecolor{chocolate}{rgb}{0.82, 0.41, 0.12}
\definecolor{cadmiumgreen}{rgb}{0.0, 0.42, 0.24}
\definecolor{ao(english)}{rgb}{0.0, 0.5, 0.0}
\definecolor{blue-violet}{rgb}{0.54, 0.17, 0.89}
\definecolor{bittersweet}{rgb}{1.0, 0.44, 0.37}

\definecolor{nightblue}{RGB}{25,25,112}
\definecolor{babyblueeyes}{rgb}{0.63, 0.79, 0.95}
\definecolor{bluebell}{rgb}{0.64, 0.64, 0.82}
\definecolor{bubblegum}{rgb}{0.99, 0.76, 0.8}
 	\definecolor{carnationpink}{rgb}{1.0, 0.65, 0.79}
\definecolor{bleudefrance}{rgb}{0.19, 0.55, 0.91}
\definecolor{blush}{rgb}{0.87, 0.36, 0.51}

\renewrobustcmd{\bfseries}{\fontseries{b}\selectfont}
\renewrobustcmd{\boldmath}{}
\newrobustcmd{\B}{\bfseries}

\newrobustcmd{\BB}{\bfseries}

%% file: format/commands_paper.tex
\definecolor{bleudefrance}{rgb}{0.19, 0.55, 0.91}
\definecolor{asparagus}{rgb}{0.53, 0.66, 0.42}
\definecolor{chocolate}{rgb}{0.82, 0.41, 0.12}
\definecolor{cadmiumgreen}{rgb}{0.0, 0.42, 0.24}
\definecolor{ao(english)}{rgb}{0.0, 0.5, 0.0}
\definecolor{blue-violet}{rgb}{0.54, 0.17, 0.89}
\definecolor{bittersweet}{rgb}{1.0, 0.44, 0.37}

\definecolor{nightblue}{RGB}{25,25,112}
\definecolor{babyblueeyes}{rgb}{0.63, 0.79, 0.95}
\definecolor{bluebell}{rgb}{0.64, 0.64, 0.82}
\definecolor{bubblegum}{rgb}{0.99, 0.76, 0.8}
 	\definecolor{carnationpink}{rgb}{1.0, 0.65, 0.79}
\definecolor{bleudefrance}{rgb}{0.19, 0.55, 0.91}
\definecolor{blush}{rgb}{0.87, 0.36, 0.51}
\definecolor{mint}{rgb}{0.24, 0.71, 0.54}
\definecolor{forest}{rgb}{0.13, 0.55, 0.13}
\definecolor{sage}{rgb}{0.74, 0.84, 0.65}
\definecolor{emerald}{rgb}{0.31, 0.78, 0.47}
\definecolor{olive}{rgb}{0.42, 0.56, 0.14}
\definecolor{jade}{rgb}{0.0, 0.66, 0.42}
\definecolor{seafoam}{rgb}{0.62, 0.91, 0.72}
\definecolor{teal}{rgb}{0.0, 0.5, 0.5}

\newcommand{\onlineappendix}{Online Appendix}

\newcommand{\thetar}[1]{$\theta_r={#1}$}
\newcommand{\thetarx}{{$\theta_r$}}

\newcommand{\MK}[1]{{#1}}

\newcommand{\ONLINE}{{{$0$-shot}}}
\newcommand{\OFFLINE}{{{$n$-shot}}}
\newcommand{\zeroshot}{{{$0$-shot}}}
\newcommand{\nshot}{{{$n$-shot}}}
\newcommand{\SQLStorm}{{{SQLStorm}}}
\newcommand{\sqlstorm}{{{SQLStorm}}}
\newcommand{\TCPDS}{{{TPC-DS}}}
\newcommand{\TPCDS}{{{TPC-DS}}}
\newcommand{\tpcds}{{{TPC-DS}}}

\newcommand{\CherryPick}{{{CherryPick}}}
\newcommand{\CherryPickAdapt}{{{CherryPick\#}}}
\newcommand{\ReLoca}{{{ReLoca}}}
\newcommand{\ReLocaAdapt}{{{ReLoca\#}}}

\newcommand{\repolink}{\url{https://github.com/mrateike/pvldb-submission}}

\newcommand{\modelperformance}{$M_{\textsc{p}}$}

\newcommand{\modelcost}{$M_{\textsc{c}}$}

\newcommand{\ScenarioA}{{Scenario $\alpha$}}
\newcommand{\ScenarioB}{{Scenario $\beta$}}
\newcommand{\ScenarioC}{{Scenario $\gamma$}}

\definecolor{ForestGreen}{RGB}{34,139,34}
\definecolor{DarkTurquoiseBlue}{RGB}{48,178,180}

\newcommand{\PartitionsperStage}{{\texttt{PartitionsperStage}}}
\newcommand{\NumberofTasks}{{\texttt{NumberofTasks}}}
\newcommand{\NumberOperations}{{\texttt{NumberOperations}}}
\newcommand{\NumberExecutors}{{\texttt{NumberExecutors}}}
\newcommand{\NumberParents}{{\texttt{NumberParents}}}

\newcommand{\ListofOperationTypes }{{\texttt{{{ListofOperationTypes}}}}}

\newcommand{\ResultSize}{{\texttt{ResultSize}}}
\newcommand{\PeakExecutionMemAllTasks}{{\texttt{PeakExecutionMemAllTasks}}}

\newcommand{\ExecutorRunTimeMax}{{\texttt{ExecutorRunTimeMax}}}

\newcommand{\ExecutorRunTimeMaxbyCore}{{\texttt{ExecutorRunTimeMaxbyCore}}}

\newcommand{\InputBytesperStage}{{\texttt{InputBytesperStage}}}

\newcommand{\DurationStage}{{\texttt{DurationStage}}}

\newcommand{\StageDuration}{{\texttt{StageDuration}}}

\newcommand{\GradientBoosting}{{\textsc{GBst}}}
\newcommand{\LinearRegression}{{\textsc{LReg}}}

\newcommand{\RandomForest}{{\textsc{RFor}}}

\newcommand{\tradeoff}{{trade-off}}
\newcommand{\Tradeoff}{{Trade-off}}

\newcommand{\nestimators}{{\texttt{n\_estimators}}}
\newcommand{\learningrate}{{\texttt{learning\_rate}}}
\newcommand{\maxdepthh}{{\texttt{max\_depth}}}
\newcommand{\subsample}{{\texttt{subsample}}}

\newcommand{\fitintercept}{{\texttt{fit\_intercept}}}

\newcommand{\bootstrap}{{\texttt{bootstrap}}}

\newcommand{\minweightfractionleaf}{{\texttt{min\_weight\_fraction\_leaf}}}

\newcommand{\Apache}{{Apache Spark\texttrademark}}
\newcommand{\Iceberg}{{Apache Iceberg\texttrademark}}
\newcommand{\Gluten}{{Apache Gluten\texttrademark}}

\newcommand{\NE}{{NE}}
\newcommand{\Section}{{\S}}
\newcommand{\Appendix}{{Appx.}}

\newcommand{\Figure}{{Figure}}

\newcommand{\Table}{{Table}}
\newcommand{\Algorithm}{{Algorithm}}

\newcommand{\Executors}{{Executors}}
\newcommand{\executors}{{executors}}
\newcommand{\executor}{{executor}}

\newcommand{\cores}{{cores}}
\newcommand{\core}{{core}}

\newcommand{\optimalne}{e^{\star}}

\usepackage{array} % Add this to your preamble

\newcolumntype{P}[1]{>{\raggedright\arraybackslash}p{#1}}

\usepackage{pdfpages}

\usepackage{atbegshi}
\AtBeginShipout{%
  \ifnum\value{page}>21
    \AtBeginShipoutDiscard
  \fi
}

%% file: sections/abstract.tex
\begin{abstract}
Allocating \executors\ (i.e. compute resources) to distributed processing systems must balance resource costs of scaling-out unnecessarily against artificial, performance-limiting bottlenecks. Naive approaches may allocate \executors\ at the application level, which have predictable costs and performance but are almost guaranteed to be sub-optimal for each of the thousands of diverse, individual stages executed by the application. 
Users may also have explicit preferences, such as completing an application within a specific time budget while minimizing cost, that existing solutions usually fail to support.
We propose a novel method for determining the number of \executors\ \emph{per stage} in a serverless \Apache\ environment, enabling users to specify their desired cost–performance trade‑off.
  Our approach trains tree-ensemble models to estimate the run times and costs of a stage as a function of allocated resources. These estimates are then used to recommend resources for each stage individually.
We evaluate our approach on \TPCDS\ and \SQLStorm\ benchmarks and compare it against two baselines.
Depending on the user-defined trade-off parameter and setup, our approach achieves $\sim$50\% cost savings across 103 \TPCDS\ queries with only a $\sim$16\% slowdown, and $\sim$40.5\% on 96 \SQLStorm\ queries at a $\sim$29\% slowdown.
\end{abstract}

%% file: sections/introduction.tex
\section{Introduction}\label{sec:introduction}

Large-scale data processing increasingly relies on data-parallel engines that execute applications on shared clusters managed by automated resource managers~\cite{kaufmann2018mira, brewer2015kubernetes,vavilapalli2013apache}.
As corporate cloud spending continues to rise organizations face increasing pressure to reduce costs without compromising performance~\cite{gartner_cloud_budgets}.
This challenge is amplified by the growing use of agentic systems where speculative querying can dramatically increase overall system workloads~\cite{liu2025supporting}.
In serverless data processing environments (e.g., serverless \Apache), billing is pay-per-use with fine-grained metering of compute and time. 
Compared to traditional pre-provisioned clusters where resource costs may be amortized or sunk, misallocation is more costly and visible in serverless settings.
A central difficulty in effective resource allocation arises from the heterogeneity of resource demands across stages that can exist within a single SQL query. 
Adding more resources to an application typically does not lead to proportional improvements in performance. 
Stage characteristics, such as number of tasks, can vary by orders of magnitude, leading to substantial differences in runtime and sensitivity to allocated resources.
Traditional, application‑level executor allocation methods assign a fixed amount of resources to each query. Every stage of the query then naively distributes tasks to the available executors.
This allocation often results in substantial misallocation as some stages become overprovisioned, increasing costs unnecessarily, while others remain under‑resourced and lead to performance slowdowns.
We address this problem by providing \emph{stage-level} optimization of executors.

A second, more practical challenge in resource allocation is that stakeholders often have specific optimization goals (e.g., minimizing runtime under a fixed cost cap.) 
Existing approaches typically optimize a single fixed system objective (e.g., execute as fast as possible) offering limited flexibility for users to express their own preferences regarding the cost–performance trade-off. 
In this work we propose a mechanism that facilitates the alignment of allocation decisions with clients' organizational objectives and budget constraints.

Our approach addresses both of the identified challenges. We propose a stage‑level, user‑preference‑aware resource allocation method. 
Although we designed it for serverless \Apache\ environments, the underlying principles, the stage-level resource allocation and explicit incorporation of user‑defined objectives can be translated to other execution platforms.
We first train simple machine‑learning models that estimate per‑stage runtime and cost as functions of the allocated executors and relevant stage features.
Their predictions are then integrated into an optimization algorithm that selects the resource configuration for each stage, guided by a user‑specified optimization objective.
Specifically, we enable users to define their own optimization goals including explicit cost–performance trade‑off parameters.
Our implementation builds on the Mira resource manager~\cite{kaufmann2018mira}, which uses stage level execution modeling to drive dynamic resource allocation with minimal scheduling overhead.

We evaluate our method on two SQL benchmarks: the widely used human-written \TPCDS~\cite{nambiar2006making} and the LLM-generated \SQLStorm~\cite{sqlstorm25}. Our evaluation compares against two state-of-the-art application-level baselines, \CherryPick~\cite{alipourfard2017cherrypick} and \ReLoca~\cite{hu2020reloca}, which we adapt to operate at the stage level for 
% a fair 
comparison. 
\MK{We refer to these adaptations as \CherryPickAdapt\ and \ReLocaAdapt, respectively.}
Depending on the user-defined trade-off parameter and setup, our approach achieves $\sim$50\% cost savings on \TPCDS\ with only a $\sim$16\% slowdown, and up to $\sim$68\% lower cost at a $\sim$49\% slowdown compared to the \Apache\ default, while also running significantly faster than \CherryPickAdapt\ at only a small cost increase.

The remainder of the paper is organized as follows. We first review background and related work (\Section~\ref{sec:background-related}), then describe the design and implementation of our framework (\Section~\ref{sec:design}), evaluate our approach empirically (\Section~\ref{sec:results}), discuss results and limitations (\Section~\ref{sec:discussion}), and conclude with a summary and outlook (\Section~\ref{sec:conclusion}).

%% file: sections/background.tex
\section{Background and Related Work} \label{sec:background-related}

\subsection{\Apache}\label{sec:spark}
\Apache\ is a distributed data processing framework for large-scale data analytics~\cite{zaharia2016apache}. 
{Applications are organized into jobs (see \Figure~\ref{fig:overview} top left).} 
When an application is submitted the framework constructs a logical Directed Acyclic Graph (DAG) based on the sequence of transformations specified by the application. This DAG is then optimized and translated into a physical execution plan comprising a series of stages~\cite{zaharia2016apache}. 
Each stage contains one or more tasks that operate on data partitions which can be executed in parallel. 
Each stage is run on a number of \executors.
\Executors\ run a fixed number of \cores\ in parallel. Each \core\ can run one task at a time. For example, one \executor\ with $4$ \cores\ can run $4$ tasks concurrently.
If the count of available \cores\ is lower than the number of tasks in a stage then \Apache\ schedules those tasks sequentially per \core.

\subsection{Mira Resource Manager}\label{sec:mira}

Resource managers allocate computational resources to applications or their components.\footnote{Allocation is distinct from scheduling, which concerns the concurrency of executed tasks, rather than the allocation of the underlying resources.} For this work, we use an advanced version of the Mira resource manager~\cite{kaufmann2018mira} for \Apache. Mira enables low-overhead resource sharing and fast scale-out across concurrently running applications on a shared cluster. However, similar to \Apache, it still relies on the task load as the key metric to make scale-out decisions: The number of allocated resources {(\executors)} doubles every second {until a predefined maximum is reached, or until the number of \cores\ equals the number of tasks} (see \Figure~\ref{fig:tradeoffs} (a)). %
While this is a simple and robust method, this scale-out strategy inherently assumes that doubling the allocated resources is a beneficial decision for performance. %As we will show later, t
This is not generally the case and creates an opportunity for novel resource allocation strategies to reduce resource costs with minimal impact on performance. In the scale-in configuration, executors are deallocated if they remain idle for 60 seconds after completing their assigned tasks.
The novel resource allocation methods proposed in this work use a modified version of Mira that limits a stage's tasks to a pre-determined number of executors.  This is in contrast to \Apache\ normal behavior which evenly distributes tasks across \emph{all} available executors.

\subsection{Machine Learning for Resource Allocation}

\CherryPick~\cite{alipourfard2017cherrypick} is a system that leverages Bayesian Optimization  to distinguish the best or close-to-the-best resource configuration (number of VMs, number of cores, RAM, among others) at a global level with a few test runs of the application.
\ReLoca\ has employed a fully connected neural network to guide the allocation of computational resources (number of \executors) by learning the impact of operations in data-parallel jobs on system overhead and execution time~\cite{hu2020reloca}. 
This approach was extended by replacing the fully connected architecture with a graph neural network, which more effectively captures the structural relationships between tasks~\cite{hu2021optimizing}.
Another approach employs a graph-based, deterministic analytical model that considers the dynamic, on-demand allocation of executors during the runtime of an \Apache\ application~\cite{tariq2023execution}. This model predicts the application's execution time by leveraging idle and backlog time metrics to evaluate executor performance.
While existing work~\cite{alipourfard2017cherrypick,ahmed2021enhanced,gao2017autopath,hu2020reloca,hu2021optimizing,tariq2023execution,venkataraman2016ernest} {has shown performance improvements in finding the optimal scale-out level of an application, it does so globally, i.e., reducing the overall job completion time for the entire application, rather than optimizing resources per application stage, as our proposed strategy does.}

Other closely related work employs Gradient Boosting and Random Forest models~\cite{sewal2022machine}. They aim to predict job execution time, stage execution time, task execution time, shuffle read/write data, and shuffle write size/records, and evaluate models using the WordCount application.
The work most closely related to ours also trains supervised ML models to predict application execution time using features available before and after execution, and evaluates these models on the test set using only pre-execution information~\cite{maros2019machine}. 
Different from us, they focus on a single TPC-DS benchmark query across varying input sizes and two other benchmarks (ML and image processing) and use a slightly different stage feature set.

Few prior works have addressed both cost and performance optimization both within~\cite{sidhanta2016optex, dimopoulos2017justice} and beyond \Apache~\cite{siddiqui2020cost, lyu2022fine}. 
However for \Apache, cost was modeled solely as a function of estimated total runtime, and optimization was limited to a single objective such as cost or fair allocation subject to a performance constraint~\cite{sidhanta2016optex, dimopoulos2017justice}. 
In contrast, we model cost and performance separately using distinct trained models, and support a range of optimization objectives that can be flexibly plugged into our framework.
We propose a \tradeoff\ algorithm, which enables users to determine the optimal number of executors to balance cost and performance according to their preferences.

%% file: sections/design-implementation.tex
\section{Design and Implementation}\label{sec:design}
We propose a two-step approach for stage-level \executors\ allocation that identifies the optimal \tradeoff\ between cost and performance, guided by a predefined objective. First, we propose two predictive models that estimate performance and cost, respectively, given a stage's features and the number of allocated \executors. In the second step, we leverage these predictions to determine the optimal number of \executors\ that balances performance and cost according to a predefined objective. 
We begin with an overview of the pipeline (\Section~\ref{sec:overview}), followed by data features and cost/performance definitions (\Section~\ref{sec:data-and-definitions}), the prediction models (\Section~\ref{sec:prediction}), and the cost-effective solver (\Section~\ref{sec:solver}).

\subsection{Overall Architecture} \label{sec:overview}

\begin{figure}
 \centering
 \includegraphics[width=\linewidth]{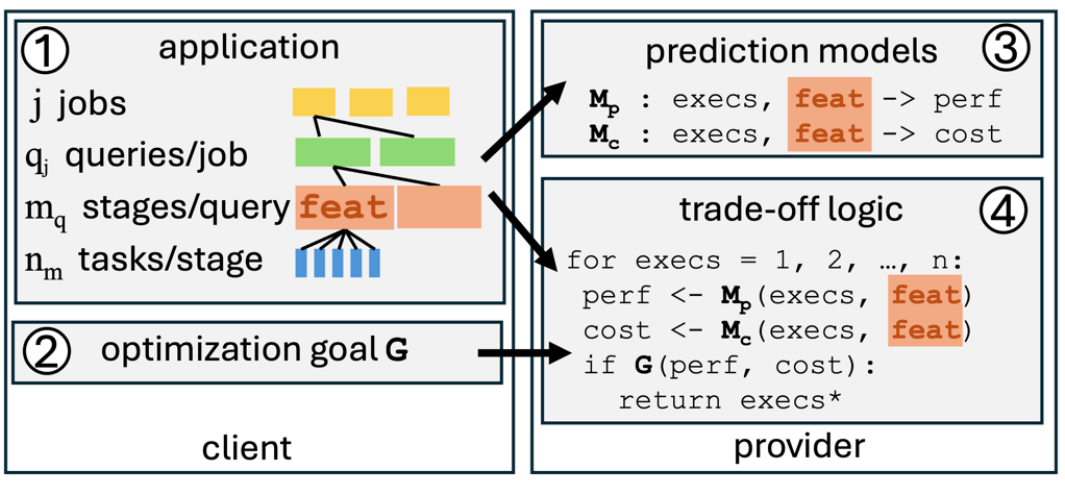}
\caption{Overview of our method (1) Given an application and (2) a client's optimization goal, we (3) learn models to predict performance and cost, which (4) we then use to determine the optimal number of \executors\ for a given stage.}
 \label{fig:overview}
\end{figure}

\Figure~\ref{fig:overview} illustrates the overall architecture of our proposed method which consists of four main components. (1) A client application (e.g., a set of SQL queries) is decomposed into jobs, each comprising stages and tasks. (2) The client specifies an optimization goal (e.g., minimizing costs while limiting performance degradation to 50\%). (3) We train prediction models that map stage features and the number of \executors\ to stage performance (\modelperformance) and cost (\modelcost). (4) A cost-effective solver uses these models and the client's objective to determine the optimal number of \executors\ for each stage, balancing performance and cost according to the client's goal.

\subsection{Data and Definitions}
\label{sec:data-and-definitions}

\paragraph{Stage-level Features.}
We collect stage-level features from Apache Spark\texttrademark{} event logs which capture stage information pre-execution, during execution, and post-execution. 
\MK{The selected features capture workload size, execution dependencies, and resource demands which are the primary factors affecting stage performance and executor allocation decisions. 
Additionally, they are independent of application semantics. 
While not the focus of this work, this feature selection may support future investigations into model generalization across diverse queries and workloads.}
Some features, such as the number of tasks, are known and logged before execution; others, such as peak memory usage, are recorded only after the stage completes (see Table~\ref{tab:features}). 
This split enables two modes of operation for our framework within the client-provider interaction.

The first and most common scenario arises when a client repeatedly executes the same application. In this case the application can be initially run $n \geq 1$ times without our optimization pipeline to collect post-execution stage-level data ($n$-shot).\footnote{We use ``$n$-shot'' to denote the setting where post-execution features are averaged over $n$ prior runs of the same application; it is not $n$-shot learning in the standard ML sense.} This data is used to (i) extract the post-execution features from the $n$ runs, record them for each stage, and use their averaged values during inference, and (ii) provide training samples that support continuous learning, improving predictions for future executions of the stage. 

The second scenario applies when a client runs an application only once or a few times. Here, repeated runs are not feasible, so predictions must rely solely on features observable prior to execution ($0$-shot).

\begin{table}[t]
\centering
\caption{Stage features available before (\ONLINE) and after (\OFFLINE) stage execution. \ListofOperationTypes\ are the different types of RDD operations, which vary across benchmarks.}
\begin{tabular}{lP{0.65\columnwidth}}
\toprule
\textbf{Availability} & \textbf{Features} \\
\midrule
\ONLINE & \PartitionsperStage, \NumberofTasks, \NumberOperations, \NumberExecutors, \NumberParents, \ListofOperationTypes \\
\midrule
\OFFLINE & 
\InputBytesperStage, \ResultSize, \PeakExecutionMemAllTasks \\
\bottomrule
\end{tabular}
\label{tab:features}
\end{table}

\paragraph{Why Standard Metrics Are Insufficient.}
A natural choice for performance is wall-clock time (\texttt{StageDuration}), and for cost the total wall-clock time charged across all allocated executors. Neither metric is well-suited to learning the relationship between executor count and stage behavior. \texttt{StageDuration} reflects (i) per-invocation overheads and cold-start latencies that cannot be attributed to specific tasks, and (ii) nondeterministic scheduling effects unrelated to the number of executors. Wall-clock cost similarly mixes scheduling artifacts with actual work performed. Below we introduce per-core time-based metrics that isolate the signal that executor allocation can directly influence.

\paragraph{Performance.}
We define the performance of a stage as the \emph{maximum executor time} $P$ (\texttt{ExecutorRunTimeMaxbyCore}), i.e., the longest duration any single core spends actively executing tasks for the stage, excluding idle time between tasks. Recall that each executor runs multiple cores in parallel. Let $\mathcal{C}$ be the set of all cores that run at least one task of a stage of interest, and let $t_c$ denote the time core $c \in \mathcal{C}$ spends running tasks of that stage. Then
\begin{equation}
    P = \max_{c \in \mathcal{C}} (t_c).
    \label{eq:performance}
\end{equation}
The intuition is that the stage's completion time is determined by the slowest core execution which is directly influenced by executor allocation. 
We treat $P$ as a time-based metric, so that maximizing performance corresponds to minimizing $P$. We generate training data such that queries do not run in parallel, which ensures that $P$ cleanly captures the per-stage workload during data collection.

\begin{figure}[t]
    \centering
    % \includesvg[width=\linewidth]{figures/correlation-matrix}
     \includegraphics[width=\linewidth]{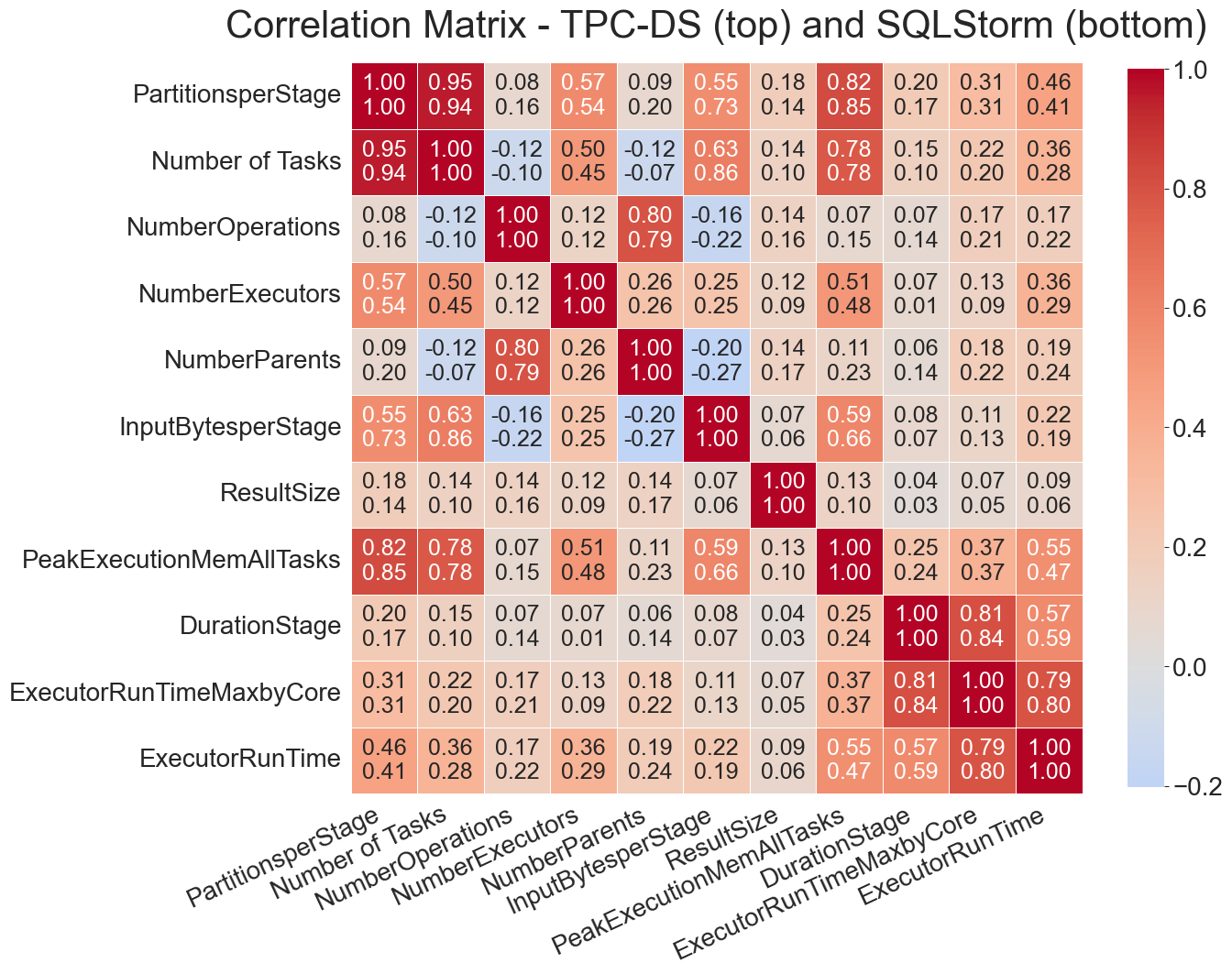}
        \caption{Feature correlation matrix for the \tpcds\ and \sqlstorm\ workload, showing pairwise Pearson correlations between model input features.}
    \label{fig:correlation-matrix}
\end{figure}

Three pieces of evidence support this choice over \DurationStage. 
First, the correlation matrix in Figure~\ref{fig:correlation-matrix} shows a strong ($> 0.8$) positive correlation between \DurationStage\ and \texttt{ExecutorRunTime-} \linebreak \texttt{MaxbyCore}, while most other features correlate more strongly with the latter, indicating it is more predictable from stage characteristics. 
Second, models trained to predict \ExecutorRunTimeMaxbyCore\ achieve lower MSE and higher $R^2$ than those predicting \texttt{Duration-}\linebreak \texttt{Stage} (see \Appendix~\ref{apx:models}). 
Third, end-to-end evaluation (\Section~\ref{sec:evaluatesolver}) shows that allocations driven by \DurationStage-based models either degrade $0$-shot runtime substantially or yield no cost reduction at $\theta_r = 0.05$ under $n$-shot, while \ExecutorRunTimeMaxbyCore-based models consistently realize the intended cost--performance trade-off. 
Since allocations made under this denoised proxy are ultimately evaluated against user-visible wall-clock time in our experiments (\Section~\ref{sec:solver}), this also confirms that decisions made under $P$ translate to real-world latency gains.

\paragraph{Cost.}
We define the cost of executing a stage as the total executor run time across all allocated executors:
\begin{equation}
    C = \sum_{\epsilon \in \mathcal{E}} t_\epsilon,
    \label{eq:cost}
\end{equation}
where $t_\epsilon$ is the run time accumulated by executor $\epsilon$ in the executor set $\mathcal{E}$. Consistent with our performance metric, $C$ is measured in core-time rather than wall-clock time, isolating the actual work performed from scheduling artifacts.

This corresponds to a linear per-core execution-time cost model, which aligns with common cloud pricing abstractions such as per-vCPU-second and per-worker-hour billing. We make this choice deliberately for clarity and tractability; the prediction and optimization steps depend only on the relative ordering of cost across configurations, so richer cost models (e.g., machine-based pricing with fixed provisioning costs~\cite{venkataraman2016ernest}, or models incorporating memory and skew effects) can be substituted without modifying the framework. We discuss this further in \Section~\ref{sec:limitations}.

Note that $C$ depends on the number of allocated executors because a task's execution time varies with its position in the scheduling order: the first task scheduled on a newly allocated core incurs spin-up overhead and therefore runs longer than subsequent tasks on the same core. For instance, Figure 3(b) allocates six tasks as first-on-core, whereas Figure 3(c) allocates only three, causing the former to incur higher overall cost despite executing the same tasks. Our methods do not change the order of tasks within stages or the order of stages within the DAG; they determine \emph{how many} executors a stage's tasks should be distributed over.

\begin{figure}
 \centering
 \includegraphics[width=\linewidth]{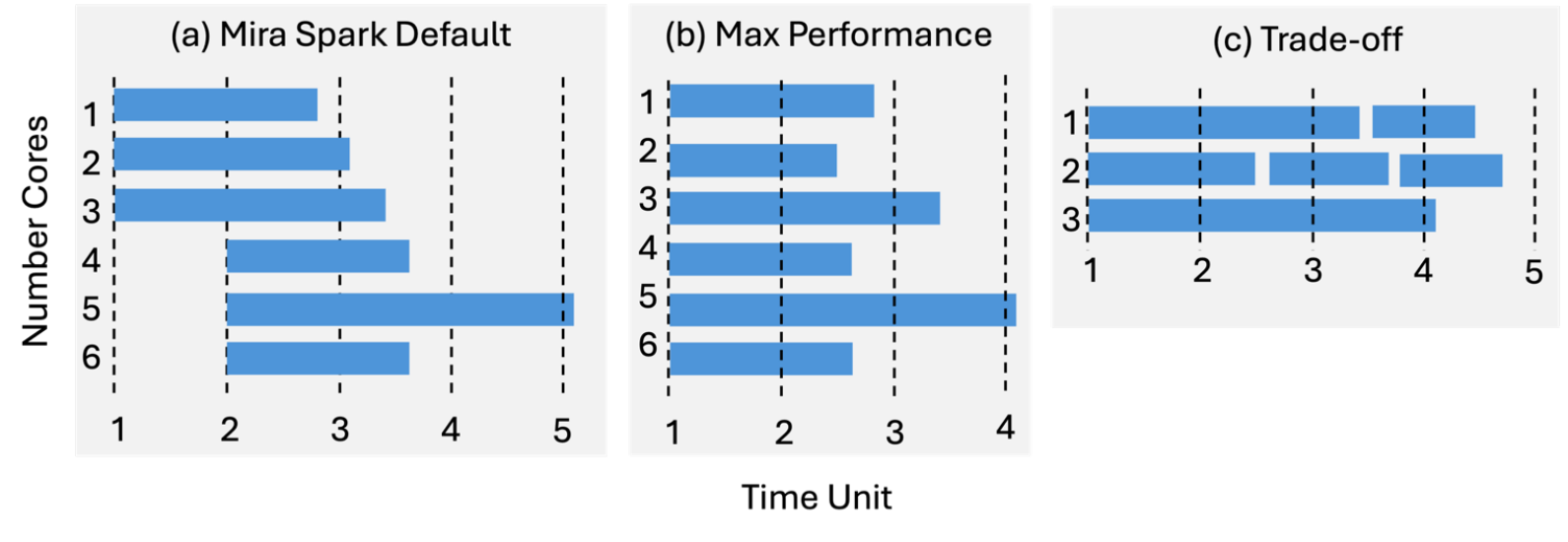}
  \caption{Illustration of different allocation strategies of a single stage with $6$ tasks (blue blocks) and $2$ \executors\ each equipped with $3$ cores: (a) {Mira Spark Default adds the second executor after a time unit} (baseline), (b) {Max Performance uses both executors from the start} (\ScenarioA), (c) {\Tradeoff\ strategies limit the tasks to a single executor based off predictive models and user preference} (\ScenarioC).}
 \label{fig:tradeoffs}
\end{figure}

\subsection{Performance and Cost Prediction}
\label{sec:prediction}

We learn two predictive models: one that estimates performance and another that estimates cost,
given a stage characterized by features $x$ and a number of \executors\ $e$.\footnote{In our implementation, we use the number of \cores\ as input, obtained directly from the Spark Environment Listener. The \tradeoff\ algorithm converts this to the number of \executors\ by dividing by the number of \cores\ per \executor.}
Although runtime and cost are highly correlated in our setting (see \Figure~\ref{fig:correlation-matrix}), separate models preserve the flexibility of the framework to accommodate alternative, potentially less correlated, cost and performance metrics.

The predicted cost and performance values are used in the second 
step (\Section~\ref{sec:solver}) to determine the optimal number of \executors\ that balances performance and cost according to the client's objective. More formally, we define the following two models:
\begin{definition}[Performance Prediction Model]
Let $x\in\mathcal{X}$ denote the feature vector of stage $i$, and let $e\in\mathbb{N}_+$ be the number of allocated \executors. The performance prediction model
\modelperformance $:\mathcal{X}\times\mathbb{N}_+\to\mathbb{R}_+$
maps $(x,e)$ to the predicted runtime
\modelperformance $(x,e)=\hat{p}_{i,e}$,
where $\hat{p}_{i,e}$ is the estimated performance measure of stage $i$ when using $e$ \executors.
\end{definition}

\begin{definition}[Cost Prediction Model]
Let $x\in\mathcal{X}$ denote the feature vector of stage $i$, and let $e\in\mathbb{N}_+$ be the number of allocated \executors. The cost prediction model
\modelcost:$\mathcal{X}\times\mathbb{N}_+\to\mathbb{R}_+$
maps $(x,e)$ to the predicted cost
\modelcost$(x,e)=\hat{c}_{i,e}$,
where $\hat{c}_{i,e}$ is the estimated cost of executing stage $i$ with $e$ \executors.
\end{definition}

 \paragraph{Interventional Training Data.}
Because we can directly control the number of \executors\ during training data generation, we fix the number of \executors\ at the beginning of each stage and keep it constant throughout the stage's execution. We then run our benchmark queries multiple times, each with a different number of \executors\ from a predefined range. This process yields data on the cost and performance of each query under varying \executor\ configurations. 
We further exclude all samples where \NumberofTasks\ is less than or equal to the number of \cores\ per \executor, since in such cases the optimal number of \executors\ is always one, eliminating the need to learn a decision-making process for such stages.

\paragraph{Non-negative Predictions.}

As discussed in the previous section, performance and cost are measured in time, which is inherently non-negative. 
Therefore, it is essential that the predictors output only non-negative values.
Off-the-shelf gradient boosting models do not enforce this constraint by default.
To address this, we explore two common approaches: (i) applying a logarithmic transformation to the target variable during training and exponentiating the predictions at inference time;\footnote{Similar to, e.g., to~\cite{siddiqui2020cost}.} and (ii) post-processing predictions by clipping negative values to a small positive threshold. 
We evaluate both methods in our experiments to assess their impact on prediction performance. Due to space constraints, we report only the results for the logarithmic transformation, which outperformed the alternative across our experiments.
% .
For the baselines, \CherryPickAdapt\  produced small single digit percentages of negative predictions, we thus apply the logarithmic transformation; \ReLocaAdapt\ yielded only non‑negative predictions while exhibiting numerical instability when using a log transform, we thus report results trained without applying the logarithmic transformation.

\subsection{Cost-effective Solver}
\label{sec:solver}

We now address how to determine the \tradeoff\ between performance (runtime) and cost when executing a stage.
Using more \executors\ speeds up execution but increases cost. Optimizing this balance improves resource efficiency and customer satisfaction, whereby the ideal \tradeoff\ depends on customer-specific goals and constraints.
We begin by presenting a non-exhaustive set of common customer scenarios (see also \Figure~\ref{fig:tradeoffs}). Then, we translate these goals into optimization problems to determine the optimal number of \executors, $\optimalne$.
Let $P(x, e)$ be the performance function and $C(x, e)$ the cost function of a stage, both determined by the stage's features $x$ and the number of allocated \executors\ $e$. $P(x, e)$ represents execution time, and minimizing it corresponds to maximizing performance.

\paragraph{\ScenarioA: Max performance.}

A client may be interested in maximizing the performance (minimize run time $P$) regardless of the costs (e.g., ``complete as fast as possible''). They thus seek for $\optimalne \in \arg\ \min \, P(x, e)$.\footnote{Alternatively, minimizing the total costs regardless of the performance decrease.}

\paragraph{\ScenarioB: Min cost subject to a performance constraint.} 
A client may aim to minimize cost while ensuring that performance does not fall below a specified threshold $\theta_P$ (e.g., ``complete in at most $\theta_P$ seconds''), seeking for
$\optimalne \in \arg \min C(x, e) \quad \text{s.t.} \quad P(x, e) \leq \theta_P$.\footnote{Alternatively, maximize performance subject to a cost constraint.}

\paragraph{\ScenarioC: Relative performance \tradeoff}

A client may seek to minimize cost up to a relative performance decrease of $\theta_r$ (e.g., ``minimize cost with at most a 50\% slowdown'') compared to the fastest configuration $e' = \arg\min_e P(x,e)$. Thus, they seek for 
$\optimalne \in \arg\min_e C(x,e) \quad \text{s.t.} \quad \frac{P(x,e) - P(x,e')}{P(x,e')} \le \theta_r$.
An absolute runtime threshold could be used instead by setting $\theta_a = (1+\theta_r)\,P(x,e')$.

\paragraph{Optimization Algorithm}
The scenarios above largely depend on both performance $P$ and cost $C$, which we predict using the models $M_p$ and $M_c$, respectively, as introduced in \Section~\ref{sec:prediction}.
To select the optimal number of \executors\ $\optimalne$, we follow a two-step approach. First, we collect predictions over a predefined list (set) of candidate executor counts $\NE$ (e.g., from $16$ to $128$)\footnote{Note, these values must fall within the distribution of the training data used to train the prediction models to ensure in-distribution prediction.}. For each stage and each executor count in $\NE$, we generate predicted performance values resulting in the list of predicted costs $\hat{C}$ and list of predicted performance values $\hat{P}$.
In the second step, we input these predictions into the optimization algorithm (see \Algorithm~\ref{alg:thresholds} for \ScenarioC\footnote{Note that in this algorithm (ln 2), we approximate for simplicity the number of \executors\ with the highest predicted cost by the number of \executors\ predicted to yield the shortest runtime, i.e., highest performance.}). The algorithm evaluates \executor\ counts in ascending order and returns the smallest count that satisfies the scenario's requirements (formulated as a stopping criterion) as the optimal number of \executors\ $\optimalne$.

\begin{algorithm}[t]
\caption{\ScenarioC: Relative performance \tradeoff. }
\label{alg:thresholds}
\begin{algorithmic}[1]
\Require Executor count list $\NE$, corresponding list of predicted performance $\hat{P}$, relative performance decrease threshold $\theta_r$. 
\Ensure Returns optimal number of \executors\ $\optimalne$.
\Function{\texttt{OPTIMALNE}}{$\NE$, $\hat{P}$, $\theta_r$}
\State $i \gets \arg\min \hat{P}$ \Comment{Index of max performance}
\For{$j \gets 0 \dots |\NE|-1$} \Comment{Iterate over executor counts}
 \If{$\frac{\hat{P}[j]-\hat{P}[i]}{\hat{P}[i]} \leq \theta_r$} \Comment{Stopping criterion}
 \State \Return $\optimalne \gets \NE[j]$
 \EndIf
\EndFor
\EndFunction
\end{algorithmic}
\end{algorithm}

%% file: sections/results.tex
\section{Evaluation}\label{sec:results}

We now evaluate our proposed framework. Note that optimality depends on the client's goal (i.e., the targeted trade-off between cost and performance), and there is no absolute ``ground truth'' for the optimal number of executors, so we cannot directly measure how close our predictions are to the true optimum.

To evaluate our framework, we first validate the cost and performance prediction models by comparing their outputs to the actual cost and runtime of a stage (\Section~\ref{sec:validateprediction}). We then evaluate the cost-effective deployment of our framework (\Section~\ref{sec:evaluatesolver}). Specifically, we test whether the cost-performance trade-off parameter and the resulting executor recommendations shift the trade-off as expected.
We begin by describing our experimental setup.

\subsection{Experimental Setup}\label{sec:experimentalsetup}

\paragraph{Benchmarks}

\TPCDS~\cite{nambiar2006making},\footnote{\url{https://www.tpc.org/tpcds/}; last accessed 27th June 2025.} is a human-written SQL query benchmark widely used in prior work~\cite{poess2007you,thanopoulou2012benchmarking,barata2015overview}. It consists of $99$ expert-curated SQL queries.
To test robustness and adaptability, we included syntactic variations of queries $14$, $23$, $24$, and $39$, which test different execution pattern with similar business logic, resulting in a total of $103$ queries.
\SQLStorm~\cite{sqlstorm25}\footnote{\url{https://github.com/SQL-Storm/SQLStorm}; last accessed 12 January 2026.} is an LLM-written benchmark which consist of complex queries designed to reflect real‑world SQL workloads. 
For our experiments, we sampled 96 queries from all queries that had sufficient runtime to collect meaningful data, specifically, those exceeding 2000 milliseconds on the deployment‑cluster hardware. \SQLStorm\ contains queries that reference three databases; we restricted our selection to those using the \TPCDS\ schema. Consequently, all sampled queries operate on the same underlying database as the \TPCDS\ benchmark queries.
This results in thousands of stage executions and provides a substantially larger evaluation workload than single-query studies often found in prior work~\cite{maros2019machine}.

Both benchmarks are executed on a dataset with $1000$ scale factor and aim to capture general characteristics of real-world decision support systems and represents the types of queries expected in client applications. Note that in practice, our method leverages client-specific or application-relevant data to train and then evaluate models on the client's application.

\paragraph{Training Data Generation and Testing Setup}

Our data processing setup uses \Apache~3.5.4 with \Iceberg\ as the table format, \Gluten\ as the execution engine, and
Mira~\cite{kaufmann2018mira} for dynamic resource provisioning.
See \Appendix~\ref{apx:hardware} for details.

We generate five replicas for training and five for testing to capture variance from \Apache\ nondeterministic scheduling, which is independent of \executor\ count.
This procedure can be thought of running the same workload multiple times with different seeds.
Each \executor\ is fixed to have $4$ cores.
For training data generation, we use $10$ static executor allocation with $8$, $12$, $16$, $24$, $32$, $48$, $64$, $80$, $96$, and $128$ executors. 

\begin{figure*}[t]
    \centering
    % -------- Row 1: Zero-shot --------
    \begin{subfigure}[t]{0.24\linewidth}
        \centering
        \includegraphics[width=\linewidth]{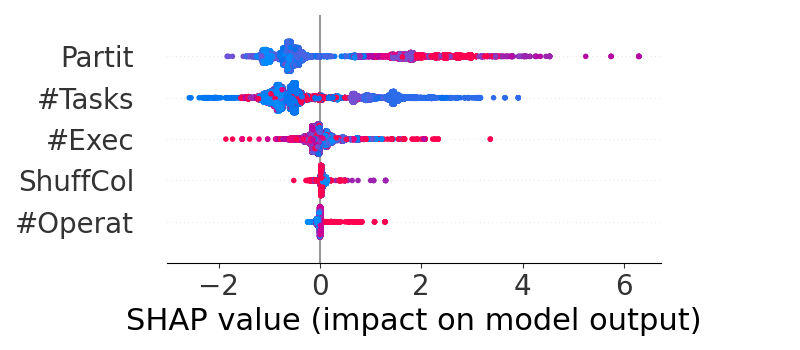}
        \caption{\zeroshot\ performance \tpcds}
    \end{subfigure}
    \hfill
    \begin{subfigure}[t]{0.24\linewidth}
        \centering
        \includegraphics[width=\linewidth]{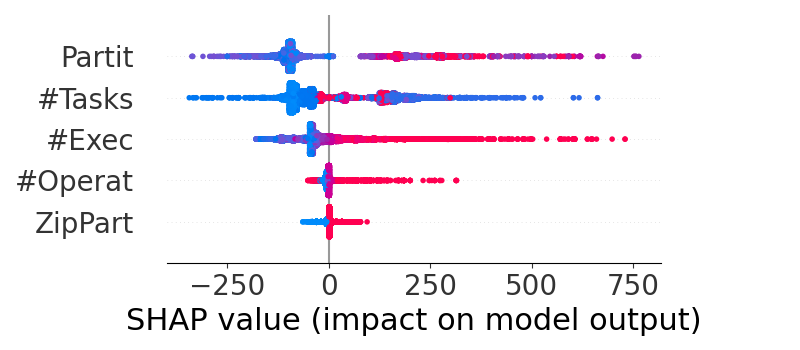}
        \caption{\zeroshot\ cost \tpcds}
    \end{subfigure}
    \hfill
    \begin{subfigure}[t]{0.24\linewidth}
        \centering
        \includegraphics[width=\linewidth]{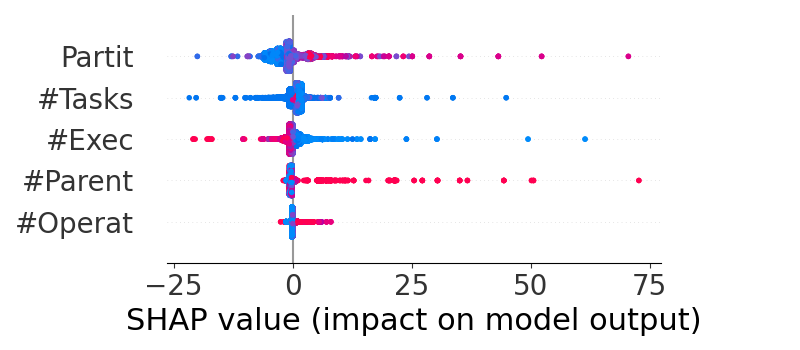}
        \caption{\zeroshot\ performance \sqlstorm}
    \end{subfigure}
    \hfill
    \begin{subfigure}[t]{0.24\linewidth}
        \centering
        \includegraphics[width=\linewidth]{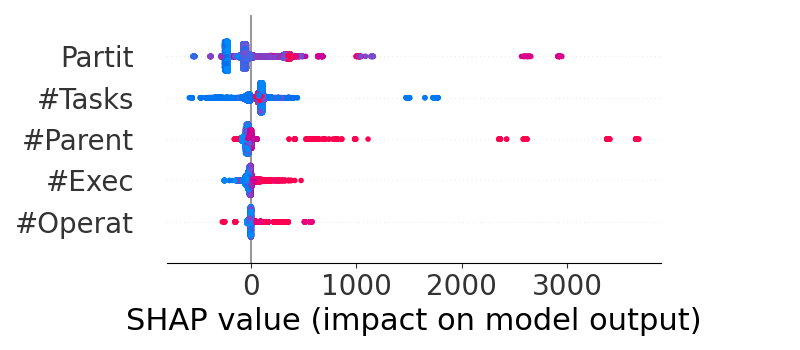}
        \caption{\zeroshot\ cost \sqlstorm}
    \end{subfigure}

    \vspace{0.5em}

    % -------- Row 2: N-shot --------
    \begin{subfigure}[t]{0.24\linewidth}
        \centering
        \includegraphics[width=\linewidth]{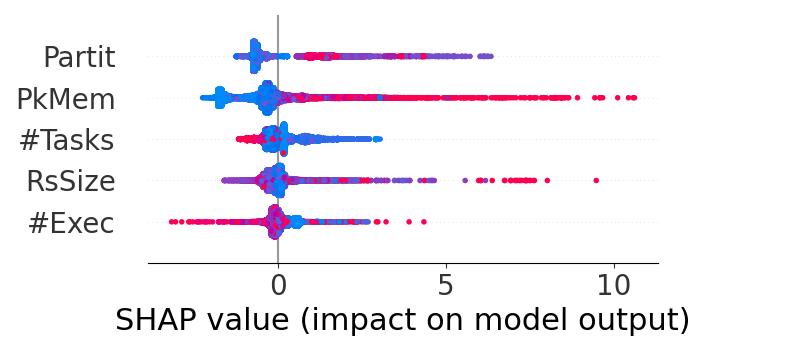}
        \caption{\nshot\ performance \tpcds}
    \end{subfigure}
    \hfill
    \begin{subfigure}[t]{0.24\linewidth}
        \centering
        \includegraphics[width=\linewidth]{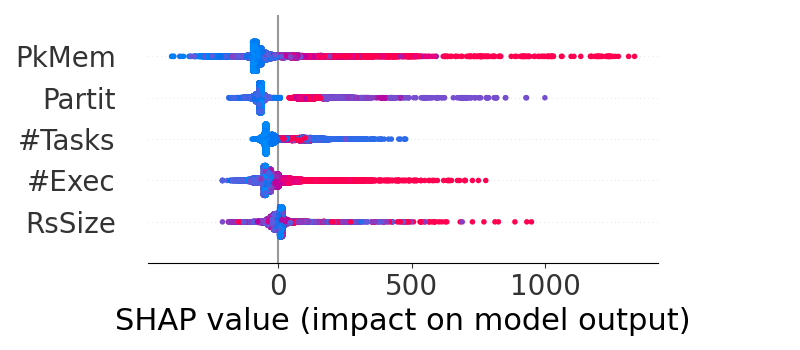}
        \caption{\nshot\ cost \tpcds}
    \end{subfigure}
    \hfill
    \begin{subfigure}[t]{0.24\linewidth}
        \centering
        \includegraphics[width=\linewidth]{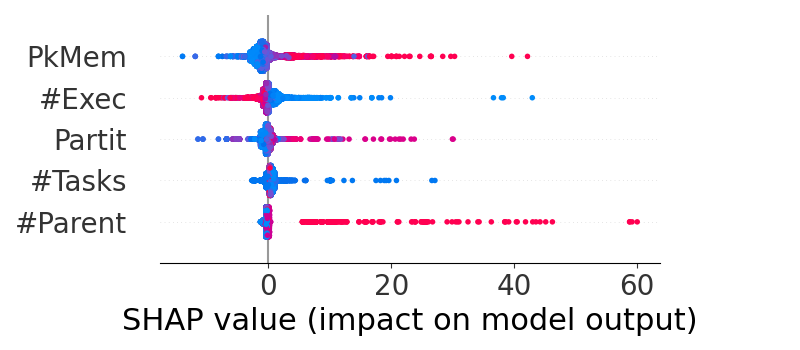}
        \caption{\nshot\ performance \sqlstorm}
    \end{subfigure}
        \hfill
    \begin{subfigure}[t]{0.24\linewidth}
        \centering
        \includegraphics[width=\linewidth]{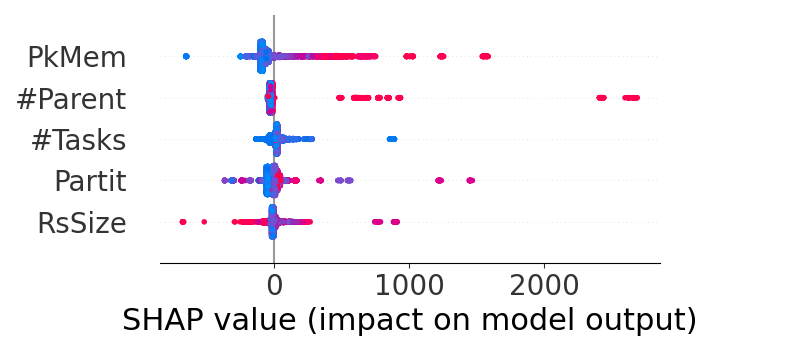}
        \caption{\nshot\ cost \sqlstorm}
    \end{subfigure}
\caption{
Top‑5 SHAP values across our models. Each point represents a stage execution; points to the right indicate that a feature increases the prediction, points to the left indicate a decrease. Red denoting higher feature values and blue lower.
Features: \PartitionsperStage\ (Partit), \NumberofTasks\ (\#Tasks), \NumberExecutors\ (\#Exec), \NumberOperations\ (\#Operat), \NumberParents\ (\#Parent), \ResultSize\ (RsSize), \PeakExecutionMemAllTasks\ (PkMem), \texttt{ShuffledColumnarBatchRDD} (Shuffcol), \texttt{WholeStageZippedPartitionsRDD} (ZipPart).
}
    \label{fig:shap-values-main}
\end{figure*}

\paragraph{Baselines.}
We compare our approach with the default Mira baselines as well as two state‑of‑the‑art methods implemented from prior work.
The default \Apache\ configuration doubles the number of executors assigned to a stage every second (see \Section~\ref{sec:mira}). 
From prior work, we adapted the implementations of Cherry- Pick
% \CherryPick
~\cite{alipourfard2017cherrypick} and \ReLoca~\cite{hu2020reloca}, which we denote as \CherryPickAdapt\ and \ReLocaAdapt, respectively.
We intentionally compare against these two methods because they are well-established optimization baselines for Spark executor configuration. We do note that our work targets a fundamentally different problem, i.e., fine-grained, stage-level cost-aware prediction, that is orthogonal and complementary to recent end-to-end or online tuning methods.
\ReLoca\ and \CherryPick\ were originally designed for application‑level prediction. We adapted their input features and modeling assumptions to our stage‑level resource‑recommendation framework. For both baselines, we use 5‑fold cross‑validation for hyperparameter tuning, consistent with our own model training. Neither method provided publicly available code, we thus reimplemented them using the information available in the respective papers. Below we summarize our implementations; further details are provided in \Appendix~\ref{apx:models} and ~\ref{apx:baselines}.

\ReLoca\ employs a neural network model (\Figure~5~\cite{hu2020reloca}). 
We adopt the original input features (\Table~1~\cite{hu2020reloca}), corresponding to our \OFFLINE\ features plus the number of cores per executor, as well as DAG width and depth metrics (\Table~4~\cite{hu2020reloca}).
\CherryPick\ is a Gaussian Process method for cost estimation used in a Bayesian Optimization framework to recommend virtual machine (VM) configuration parameter setups.
We modified it to take as input our \OFFLINE\ features and train their Gaussian Process model to predict cost and performance.
For end-to-end testing, we allocate executors according to Algorithm~\ref{alg:thresholds} with \thetar{0.0}, matching our setup.
Although the baselines are adapted as carefully and fairly as possible, differences in original prediction granularity and input feature requirements may affect direct comparability.

Our method and all baselines are trained on the same interventional dataset, ensuring a fair comparison. 
The training data consists of 10 executor configurations with 5 repetitions each (50 runs per benchmark).
We note that this dataset size is substantially smaller than the approximately 3,000 runs used by \ReLoca\ and the 66 cloud configurations evaluated by \CherryPick, which may contribute to higher prediction error for the neural network based baselines.
This, however, highlights a key advantage of our approach, which can achieve effective resource allocation with significantly less training data.

\paragraph{Technical Specifications}
The data processing infrastructure leverages Apache Spark 3.5.4 integrated with Iceberg as the underlying table format. 
Iceberg is configured with a Hive-backed catalog and an external Hive Metastore for metadata management while table data is stored in S3-compatible object storage accessed via the S3A connector. 
The execution engine is accelerated using Gluten with the Velox native backend, enabling columnar execution and native vectorized processing.

The maximum available number of executors is 128. 
For Training, we shut down an executor after 60 seconds of idleness, following the default \Apache\ configuration. 
For testing, Apache Spark queries Mira for updates on the recommended number of cores computed by our optimization framework at 300‑millisecond intervals. 
Multiple stages may run in parallel as long as their combined recommended executors remain within the executor limit. 
For our methods and \CherryPickAdapt, any executor that becomes idle is shut down immediately when no further tasks remain for its active stage. 
The Apache Spark baseline uses the default 60‑second executor idle timeout.

\subsection{Validation of Prediction Models}\label{sec:validateprediction}

We first evaluate the test performance of our models and baselines using 5‑fold cross‑validation on the interventional training data. We then assess, in the following section (\Section~\ref{sec:evaluatesolver}), their end‑to‑end performance when used to predict and adjust resource allocations on benchmark runs.
Importantly, the evaluation of the test performance should be interpreted in context: the primary objective of our proposed framework is not highly precise runtime prediction, but rather the effective use of these predictions to support resource allocation decisions.
Theoretical work has indeed shown that near‑optimal allocation decisions can be achieved despite imperfect or inaccurate predictions~\cite{casacuberta2026good}.

\input{tables/table2}

\paragraph{Metrics.} We evaluate the prediction performance of the models using Mean Squared Error (MSE) and $R^2$ score. MSE quantifies the average squared difference between the actual and predicted values. A lower MSE indicates better model performance. However, its absolute value is difficult to interpret, as it is highly dependent on the scale of the target variable. $R^2$ gives an indication of how well the model explains the variance in the target variable. A higher $R^2$ suggests that the model captures a significant proportion of the variability of the data. 
{An $R^2$ value between 0 and 1 indicates the proportion of variance in the target variable explained by the model.}

\paragraph{Our Models.}
For our approach, we evaluate several standard Scikit‑learn~\cite{scikit-learn} models, including penalized linear regression \linebreak (Lasso, Ridge), Random Forests, and Gradient Boosting (GB). 
Linear regression and Random Forests consistently underperformed on the test set compared to GB. 
This finding is consistent with prior work showing that tree-based models perform best for predicting \Apache\ performance across different configuration settings~\cite{wang2016novel}.
We omit these results in the main paper and report only the GB configuration. Details on the hyperparameter search space and selected models are provided in \Appendix~\ref{apx:models}.

\paragraph{Results.}
\Table~\ref{tab:modelselection} reports the validation performance of our models and baselines in predicting stage-level performance and cost.
We first observe that \ReLocaAdapt\ attains substantially higher MSE and lower $R^2$ scores ($0.1574$–$0.3352$) across benchmarks, i.e., fails to predict cost and performance behavior reliably.
Due to its poor predictive performance, we omit \ReLocaAdapt\ from subsequent runtime evaluations (\Section~\ref{sec:evaluatesolver}).
In contrast, \CherryPickAdapt\ performs competitively with our methods. Although its inference inputs, like those of \OFFLINE, depend on post-execution features, its MSE remains closer to our \ONLINE\ (and higher than \OFFLINE), while its $R^2$ is more aligned with (and often a slightly lower than) \OFFLINE.
Our \OFFLINE\ model consistently achieves the lowest MSE and highest $R^2$ ($0.9186$–$0.9962$), outperforming all other models.
Overall, these results demonstrate that our models, particularly the \OFFLINE\ model, provide highly reliable predictions of both cost and performance.

\Figure~\ref{fig:shap-values-main} reports the top‑5 SHAP values~\cite{SHAP_NeurIPS,lundberg2020local2global}, which quantify each feature's contribution to the model's predictions by indicating both its relative importance and whether it increases or decreases the predicted value. 
More results in \Appendix~\ref{apx:results}.
We make the following observations:
For \zeroshot\ on both datasets, the most influential features for cost and performance prediction are 
\texttt{Partitionsper-} \texttt{Stage},
\NumberExecutors, \NumberofTasks, and \texttt{NumberOpera-} \linebreak \texttt{tions}, 
indicating that parallelism, task granularity, and overall plan size are dominant predictors.
For \tpcds, we additionally observe the importance of specific Spark execution operators. 
In performance prediction, the presence of \texttt{ShuffledColumnarBatchRDD} is particularly influential, while in cost prediction it is \texttt{WholeStage-} \texttt{ZippedPartitionsRDD}. 
This suggests that low‑level execution \linebreak mechanisms related to columnar shuffling, and zipping multiple input RDD partitions together are critical drivers of model predictions on this benchmark.
In contrast, for \sqlstorm, \NumberParents\ strongly influences cost and performance prediction, suggesting a larger importance of execution‑plan structure and dependencies.
For \nshot\ on both datasets, the most influential features for cost and performance prediction are 
\PartitionsperStage, 
\PeakExecutionMemAllTasks, \NumberExecutors, and \texttt{Number-} \linebreak \texttt{ofTasks}. 
For \tpcds, \ResultSize\ additionally appears among the most influential features. 
In \sqlstorm, \NumberParents\ again emerges as an important predictor for performance, while for cost prediction \NumberExecutors\ does not ranks among the top-5 features; instead, \ResultSize\ and \NumberParents\ appear in the top five most influential predictors.
These results can be explained by differences in benchmark design. \sqlstorm\ is intentionally constructed to generate a wide range of queries, from simple to complex, resulting in substantially greater diversity in query plans and execution traces than \tpcds~\cite{sqlstorm25}.
Consequently, \tpcds\ is more operator‑driven, whereas \sqlstorm\ is more structure‑driven, explaining why execution‑plan depth and dependency complexity (\NumberParents) emerge as key predictors of both cost and performance.

For \NumberExecutors\ in \tpcds, higher executor counts lead to lower predicted runtimes (better performance), while fewer executors result in higher predicted costs. This is matching general expectations. In contrast, for \sqlstorm, larger numbers of executors are generally associated with higher predicted runtimes and costs. An exception is the \nshot\ setting, where the effect of \NumberExecutors\ is more mixed, contributing to both lower runtimes and higher costs.
Importantly, the role of the SHAP analysis in our framework should be viewed in context: the models are primarily intended to inform allocation decisions rather than to deliver highly precise runtime predictions. 
Accordingly, SHAP values are informative only insofar as they provide actionable insights that support these allocation decisions.

\subsection{Evaluation of Cost-effective Solver Deployment}\label{sec:evaluatesolver}

\begin{figure}[t]
  \centering

  % --- Top subfigure: TPC-DS ---
  \begin{subfigure}{\linewidth}
    \centering
    \includegraphics[width=\linewidth]{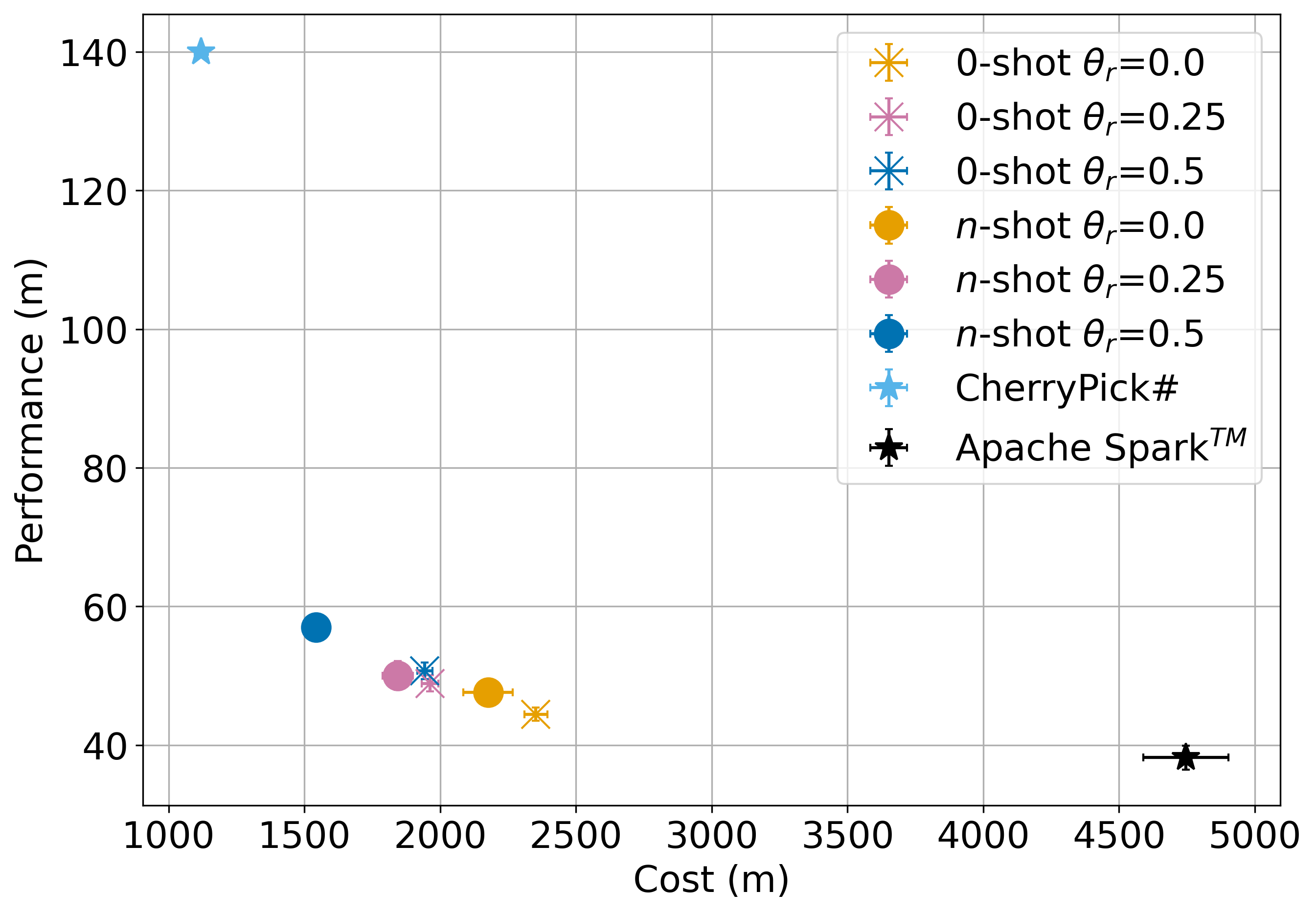}
    \caption{\TPCDS}
    \label{fig:perfvscost_tpcds}
  \end{subfigure}

  \medskip

  \begin{subfigure}{\linewidth}
    \centering
    \includegraphics[width=\linewidth]{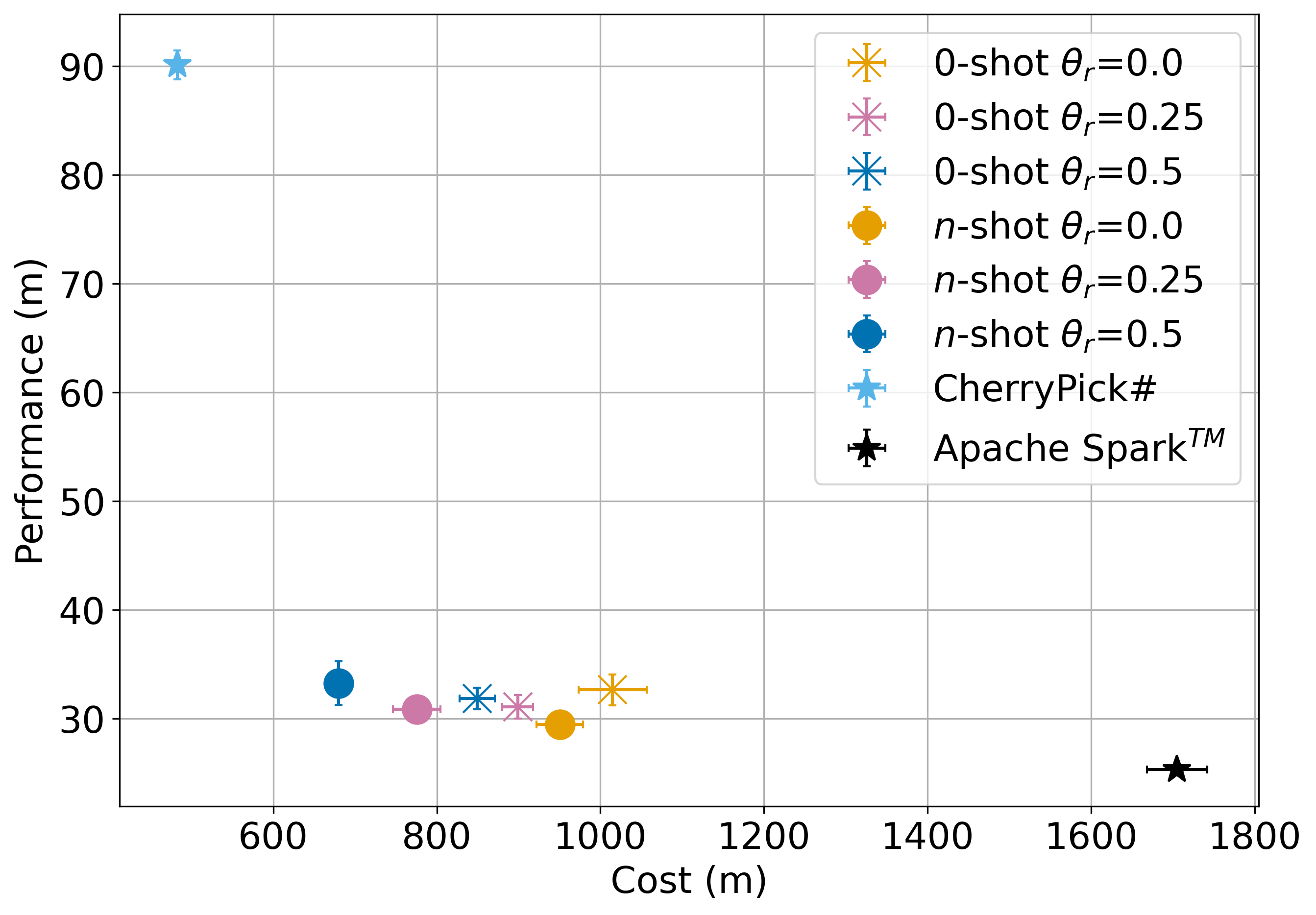}
    \caption{\SQLStorm}
    \label{fig:perfvscost_sqlstorm}
  \end{subfigure}
\caption{Cost–performance \tradeoff\ for \ScenarioC\ across benchmarks, our \ONLINE\ and \OFFLINE\ methods and baselines (\Apache, \CherryPickAdapt). Cost ($\downarrow$) and performance ($\downarrow$) in minutes over 5 runs.}
\vspace{-5pt}
  \vspace{-5pt}
  \label{fig:perfvscost_overall}
\end{figure}

\begin{figure}[t]
    \centering
    \includegraphics[width=\linewidth]{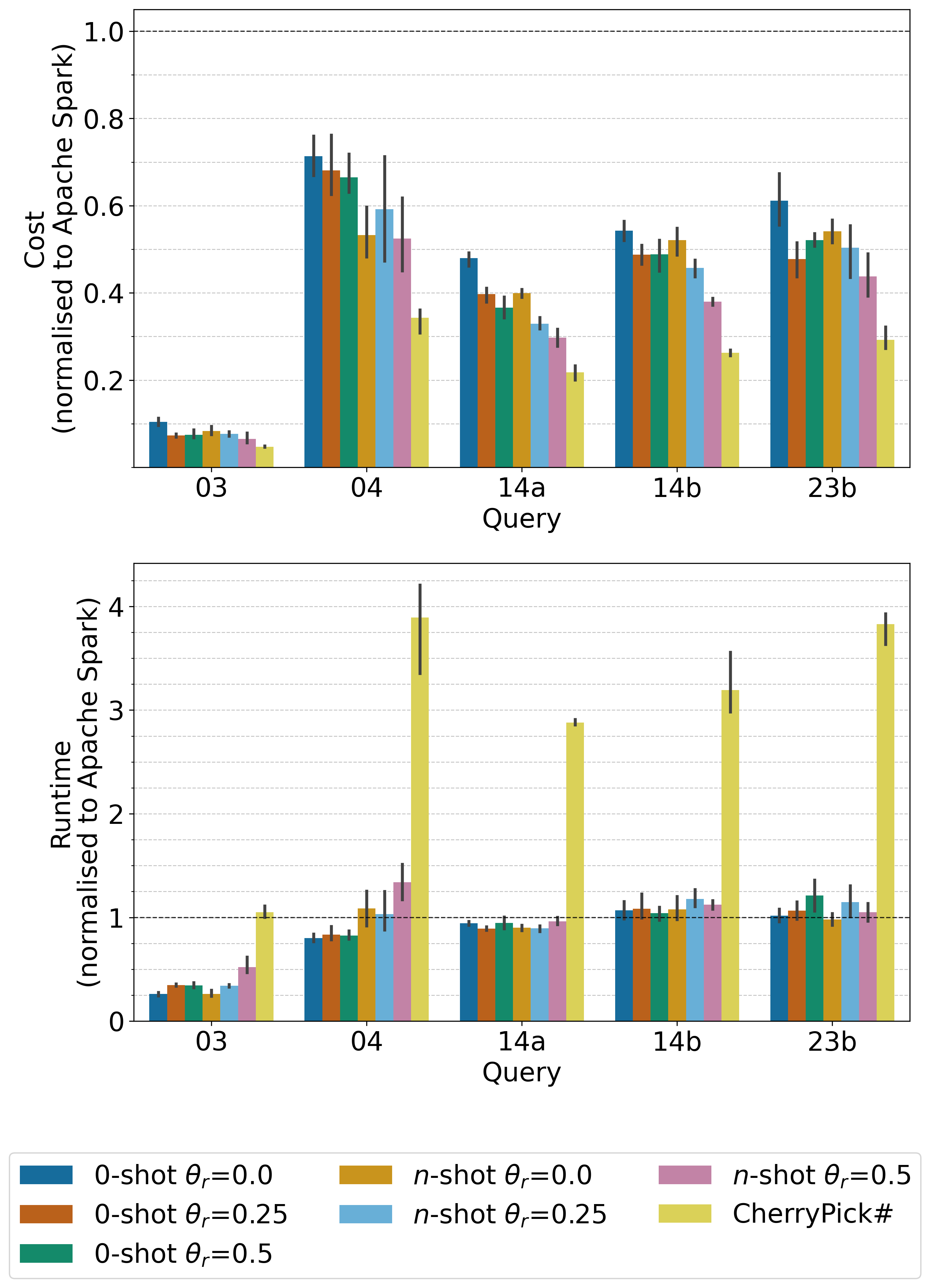}
    \caption{Normalized cost (top) and execution time (bottom) for the five \TPCDS\
queries with the largest absolute cost difference between \Apache\ and our \ONLINE\ $\theta_r=0.0$.
Showing mean values ($\pm 1$ standard deviation error bars) across $5$ runs, normalized by \Apache\ mean for the same query (dashed line at~$1.0$); values below~$1.0$ indicate improvement over the baseline.}
    \label{fig:costperftop}
    \vspace{-5pt}
\end{figure}

We now evaluate our cost-effective solver in a real-world setting.
We focus on \ScenarioC\ (see \Section~\ref{sec:solver}), which we consider the most practical, and run Algorithm~\ref{alg:thresholds} with varying values of $\theta_r$.
 This user-supplied parameter can be thought of as a throttle on the performance of the stage, with $\theta_r = 0.0$ being no throttle (full performance) and $\theta_r > 0$ reducing the recommended resources. 
We compare the cost and performance of running both benchmarks using our recommendation algorithm with \OFFLINE\ and \ONLINE\ models against Mira \Apache's default resource allocator~\cite{kaufmann2018mira}, which doubles resources every second until demand is met, as well as the \CherryPickAdapt\ baseline.
The evaluation is end‑to‑end: we report cost and performance metrics aggregated over the execution of the entire benchmark across all queries, reflecting net benefits in overall latency and cost.

\paragraph{End-to-End Results.}
% %
\Figure~\ref{fig:perfvscost_overall} illustrates how different resource‑ allocation strategies influence both cost (as defined in \Section~\ref{sec:data-and-definitions}) and performance (importantly, measured here as the wall‑clock time from query initiation to completion) for both benchmarks (\TPCDS, \SQLStorm).
Metrics are aggregated over all benchmark queries.
% %

First, across both benchmarks, we observe that \Apache's default allocation method achieves the fastest execution times while it also incurs the highest costs, whereas \CherryPickAdapt\ yields the lowest costs while also exhibiting the slowest execution times (lowest performance). Our \ONLINE\ and \OFFLINE\ methods show a balance between these two extremes, while achieving different performance–cost trade‑offs depending on the value of the $\theta_r$ parameter. 

Compared to \Apache, our methods deliver substantial cost reductions while maintaining competitive performance. 
We achieve a cost reduction of at least $\sim$50.4\% on \TPCDS\ and $\sim$40.5\% on \SQLStorm\ at an increase of runtime by $\sim$16.4\% and $\sim$29\%, respectively (\ONLINE\  \thetar{0.0}), and cost reductions up to $\sim$67.5\% on \TPCDS\ and $\sim$60.1\% on \SQLStorm, with an increase of run time of $\sim$49.3\% and $\sim$31.4\%, respectively (\OFFLINE\ \thetar{0.5}).
Compared to \CherryPickAdapt, our methods show substantial performance improvements while incurring modest cost increases. Especially for \OFFLINE\ \thetar{0.5}, our method runs \TPCDS\ 59.3\% faster at an 37.8\% cost increase, and 63.1\% faster on \SQLStorm\ at an 40.9\% cost increase. 
These results demonstrate that our methods, especially \OFFLINE, achieve a better performance-cost trade-off compared to both baselines across both benchmarks.

The trade-off parameter \thetarx\ serves as a control mechanism for balancing performance and cost objectives in resource allocation. Increasing \thetarx\ reduces costs across both benchmarks and methods. On \TPCDS, increasing \thetarx\ from $0.0$ to $0.5$ reduces costs from $2353$ to $1943$ mins for \ONLINE\ and from $2176$ to $1544$ mins for \OFFLINE. A similar trend can be observed for \SQLStorm. 
For \TPCDS, this cost reduction comes with the expected increase in runtime across models, whereas for \SQLStorm\ \ONLINE, $\theta_r>0$ reduces both cost and runtime compared to \thetar{0.0}.

\ONLINE\ and \OFFLINE\ differ in the data available at inference time. The \ONLINE\ method operates based on stage characteristics known before execution, while \OFFLINE\ additionally leverages stage information available after execution that was collected over the past $n$ runs. For the same value of \thetarx, \OFFLINE\ consistently achieves across benchmarks lower cost than \ONLINE, and in some achieves Pareto‑optimality (both lower cost and lower runtime). We attribute this \OFFLINE\ performance to its more accurate cost predictions enabled by the richer input data.

\paragraph{Query-level Results.}
\Figure~\ref{fig:costperftop}
presents normalized cost (top) and performance (bottom) for five \TPCDS\ queries
that showed the greatest divergence in cost between our \ONLINE\ (\thetar{0.0}) approach and the \Apache\ baseline.
Cost and performance metrics are normalized per query to the
Apache~Spark mean (Apache~Spark $= 1.0$); values below 1 indicate
improvement over the baseline.
For results across all queries, see~\Appendix~\ref{apx:results}.
Query~3 sums discount amounts for items from a specific manufacturer sold at a specific time, grouped by year and brand, returning the top results.
Query~4 computes yearly customer sales across channels, and returns customers whose catalog sales growth exceeds both store and web growth.
Queries~14 (a/b) select items sold across channels over a given period, and compute sales and counts by channel, brand, class, and category for groups exceeding average sales.
Query~23 (a/b) sums sales from different channels for a given period for frequently sold items and high-spending customers.

We observe that \ONLINE\ and \OFFLINE\ variants of our method behave similarly across individual queries, while there is a large difference to
\CherryPickAdapt\ baselines.
For query 3, our approaches reduce costs to below 10\% of \Apache's, whereas \CherryPickAdapt\ lowers costs further to about 5\%. However, this additional cost reduction comes with execution times slightly above \Apache, while our methods achieve execution times of only 0.25-0.5\% of \Apache's.
% %
The displayed queries appear to demand substantially greater resources from \Apache, which, as mentioned above, doubles its allocated resources at each time step until a limit is reached.
This suggests that the advantages of our methods arise primarily from their improved handling of more resource‑intensive queries.

\begin{figure}[t]
 \centering
 \includegraphics[width=\linewidth]{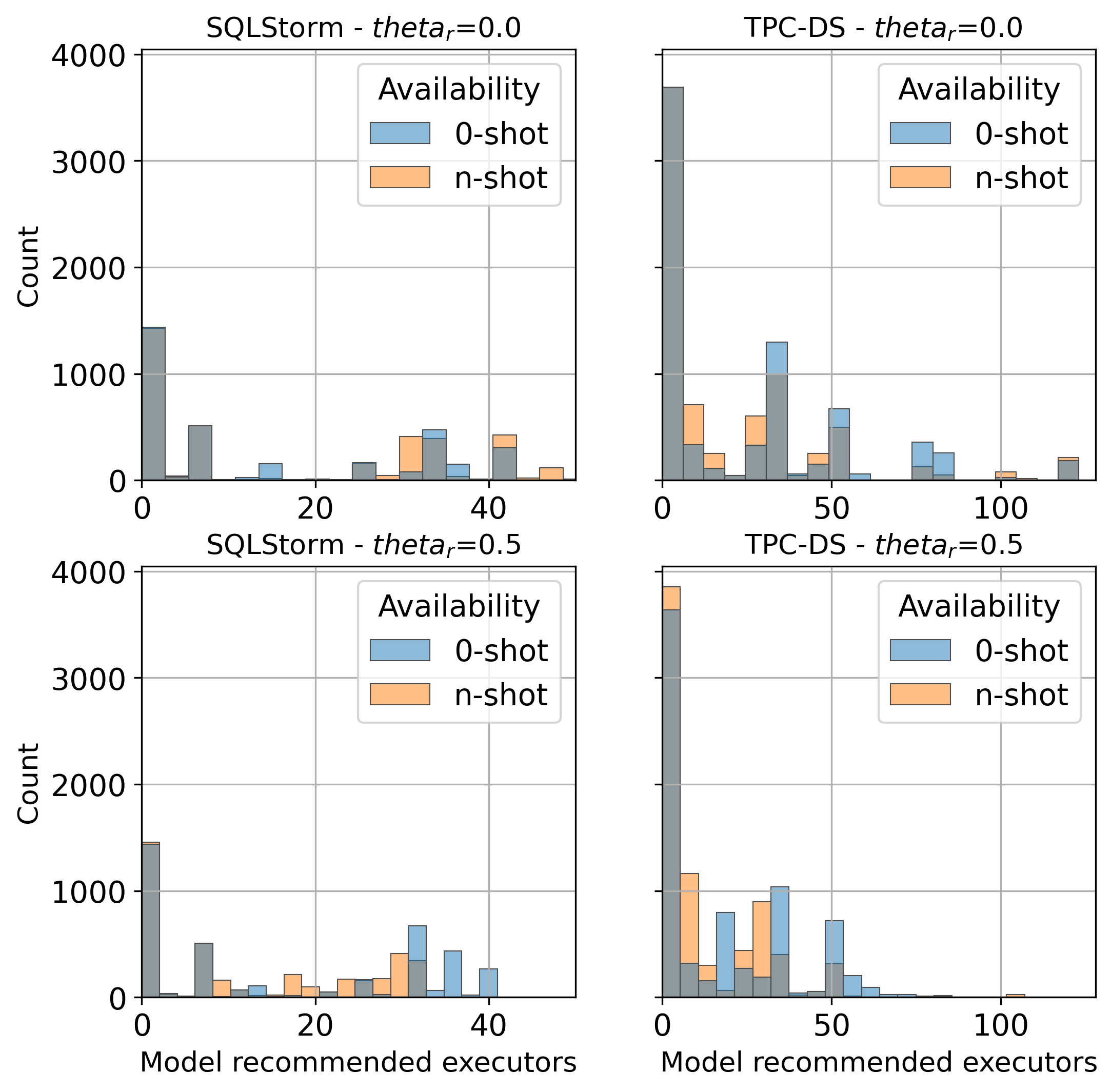}
 \caption{Distribution of recommended \executors\ over all stages of all queries for \ONLINE\ (left) and \OFFLINE\ (right) with $\theta_r=0.0$ (top) and $\theta_r=0.5$ (bottom); \ScenarioC. Across 5 runs.} % 
 \label{fig:recommendedcores}
\end{figure}

\paragraph{Investigation of Executor Allocations.}
\Figure~\ref{fig:recommendedcores} shows the distribution of recommended \executors\ for our \ONLINE\ and \OFFLINE\ methods across both benchmarks for \thetar{0.0} and \thetar{0.5}. 
This provides insight into how the different methods realize the desired cost-performance tradeoff and whether they achieve it through substantially different resource provisioning strategies.
We remind the reader that selecting between \OFFLINE\ and \ONLINE\ modes, and tuning \thetarx, depends on user- and application-specific priorities. 

Overall, the recommended number of executors for \SQLStorm\ ranges from 1-54, and for \TPCDS\ ranges from 1-123.
At \thetar{0.0}, we observe a wider spread of recommendations compared to \thetar{0.5} across benchmarks and models, reflecting that the stronger cost‑reduction objective for \thetar{0.5} favors smaller number of \executors. 
Across both benchmarks, the \OFFLINE\ method recommends on average less executors than \ONLINE at a lower variance.

%% file: tables/table2.tex
\begin{table}[]
\setlength{\tabcolsep}{3.7pt} % 
 \centering
  \caption{Model performance for different benchmarks (B) and targets (T). Cost and Performance (Perf.), Dataset (D), Mean Squared Error (MSE, in seconds) and $\mathbf{R^2 \times 10}$. Results averaged over 5-fold cross-validation, our methods marked *. Best (significant) bold.} 
 \begin{tabular}{lllrl}
 \toprule
B & T & Model & MSE ($\downarrow$) & R$^2$ ($\uparrow$) \\
 \midrule
\multirow{8}{1em}{\rotatebox{90}{\TCPDS}}  &  \multirow{4}{1em}{\rotatebox{90}{Perf.}} & \ReLocaAdapt & $\phantom{0}27.11 \pm \phantom{0}8.41$ & $33.52 \pm 12.84$\\
 &   & \CherryPickAdapt & $\phantom{0}3.34 \pm \phantom{0}0.43$ & $89.28 \pm \phantom{0}4.06$ \\
 &  & \ONLINE* &  $\phantom{0}2.90 \pm \phantom{0}0.18$ & $54.34 \pm \phantom{0}1.56$ \\
&   & \OFFLINE* &  $\mathbf{\phantom{0}0.51 \pm \phantom{0}0.05}$ & $\mathbf{91.86 \pm \phantom{0}0.76}$  \\
 \cline{2-5}
&  \multirow{4}{1em}{\rotatebox{90}{Cost}} & \ReLocaAdapt & $170,842.84 \pm 30,056.95$ & $19.04 \pm 12.02$ \\
&    & \CherryPickAdapt &  $\phantom{0}36,859.59 \pm \phantom{0}4,823.53$ & $81.08 \pm \phantom{0}2.51$  \\

&  & \ONLINE* &  $\phantom{0}38,652.09 \pm \phantom{0}3,158.79$ & $67.66 \pm \phantom{0}1.49$\\
&   & \OFFLINE* &  $\mathbf{\phantom{000}4,592.48 \pm \phantom{000}334.10}$ & $\mathbf{96.14 \pm \phantom{0}0.35}$  \\
\midrule
\multirow{8}{1em}{\rotatebox{90}{\SQLStorm}}  &  \multirow{4}{1em}{\rotatebox{90}{Perf.}} & \ReLocaAdapt & $\phantom{0}40.77 \pm 11.79$ & $15.74 \pm \phantom{0}2.43$  \\
 &   & \CherryPickAdapt &  $\phantom{0}3.33 \pm \phantom{0}0.56$ & $92.70 \pm \phantom{0}1.78$  \\
 &  & \ONLINE* &  $\phantom{0}2.67 \pm \phantom{0}0.20$ & $94.00 \pm \phantom{0}1.76$ 	\\
&   & \OFFLINE* &  $\mathbf{\phantom{0}0.65 \pm \phantom{0}0.50}$ & $\mathbf{98.78 \pm \phantom{0}0.61}$  \\
 \cline{2-5}
&  \multirow{4}{1em}{\rotatebox{90}{Cost}} & \ReLocaAdapt & $188,705.44 \pm 48,227.33$ & $20.07 \pm 13.63$  \\
&    & \CherryPickAdapt &  $\phantom{0}25,239.84 \pm \phantom{0}6,276.70$ & $89.35 \pm \phantom{0}1.99$  \\
&  & \ONLINE* &  $\phantom{0}11,545.14 \pm \phantom{000}358.13$ & $94.98 \pm \phantom{0}0.75$ 	\\
&   & \OFFLINE* &  $\mathbf{\phantom{000}856.26 \pm \phantom{000}522.65}$ & $\mathbf{99.62 \pm \phantom{0}0.25}$ \\
 \bottomrule
 \end{tabular} 
 \label{tab:modelselection}
\end{table}

%% file: sections/discussion.tex
\section{Discussion and Limitations}
\label{sec:discussion}

\subsection{Practical Insights}
\label{sec:insights}

Our framework and experimental results yield three practical insights for resource allocation in serverless data processing.

\paragraph{Allocation Quality Does Not Require Prediction Accuracy.}
Although we frame the method in terms of predicting per-stage runtime and cost, our results show that allocation quality depends on preserving the \emph{relative} cost-performance ordering of stages across resource configurations, not on the absolute accuracy of either prediction. This is consistent with recent theoretical results showing that near-optimal allocation can be achieved with substantially fewer samples than accurate prediction requires~\cite{casacuberta2026good}. In practice, this means a small interventional training set (50 runs per benchmark in our setup) suffices to drive effective allocation, which is an order of magnitude less than the ${\sim}3{,}000$ runs used by ReLoca~\cite{hu2020reloca} or the 66 cloud configurations explored by CherryPick~\cite{alipourfard2017cherrypick}.

\paragraph{Stage-Level Allocation Closes a Structural Gap in \Apache's Scale-Out Heuristic.}
\Apache's default strategy doubles allocated executors every second until task demand is met. This eager scale-out systematically over-provisions stages whose tasks finish quickly, where the marginal executor contributes little but is still billed. Our stage-level formulation reasons explicitly about the number of executors each stage should receive, and the resulting allocations (\Figure~\ref{fig:recommendedcores}) concentrate around far smaller executor counts than \Apache\ would have allocated, while recovering most of \Apache's runtime. On \tpcds, this translates to ${\sim}50\%$ lower cost at a ${\sim}16\%$ runtime increase relative to the Spark default ($0$-shot, $\theta_r = 0.0$).

\paragraph{User-Defined Trade-Offs Are the Right Interface for Serverless Allocation.}
Users of pay-per-use environments typically express preferences as objectives (``minimize cost subject to a runtime budget'') rather than as resource plans. Our $\theta_r$ parameter exposes the cost-performance frontier directly: increasing $\theta_r$ from $0.0$ to $0.5$ moves the operating point smoothly along this frontier on both benchmarks, allowing users to select allocations that match organizational priorities. This is, to our knowledge, the first stage-level allocation method that exposes such an interface.

\subsection{Limitations}
\label{sec:limitations}

\paragraph{Cost Model.}
We instantiate the framework with a linear per-core execution-time cost model, which aligns with common cloud pricing abstractions such as per-vCPU-second and per-worker-hour billing. The framework itself depends only on relative cost ordering across configurations and is compatible with richer cost models (e.g., machine-based pricing with fixed provisioning costs~\cite{venkataraman2016ernest}, or models incorporating memory and skew effects). Integrating and evaluating such models is an important direction for future work.

\paragraph{Prediction Model Family.}
We adopt tree-based ensemble models, which capture non-linear feature interactions well and provide stable predictions within the training distribution. As is common for learned system models~\cite{wang2016novel}, predictive accuracy degrades under substantial distribution shift, for instance when executor counts or hardware configurations lie far outside the training range. We view this as an acceptable trade-off given that production cluster configurations and workloads tend to be stable over meaningful time horizons, and that incremental retraining can absorb gradual drift.

\paragraph{Evaluation Scope.}
Due to confidentiality constraints, our evaluation relies on two public benchmarks rather than production workloads. \tpcds\ and \sqlstorm\ together span thousands of stage executions and a broad spectrum of analytical query patterns, from human-curated decision-support queries to LLM-generated complex queries, but they cannot fully represent the heterogeneity of production environments. Evaluation is further bounded by the executor configurations and scale factor used during data collection.

\subsection{Deployment Considerations}
\label{sec:deployment}

\paragraph{When to Use $0$-shot vs.\ $n$-shot.}
The two modes target different deployment regimes. The $n$-shot mode achieves consistently better cost--performance trade-offs (\Figure~\ref{fig:perfvscost_overall}) and is the natural choice when an application is executed repeatedly, since the post-execution features it relies on can be extracted from prior runs. The $0$-shot mode applies when only pre-execution features are available, such as for one-off queries or newly deployed applications, and still delivers substantial cost reductions, though with somewhat higher variance in recommended executor counts (\Figure~\ref{fig:recommendedcores}).

\paragraph{Cost of Training Data Collection.}
Generating the interventional training set is a one-time, offline cost amortized over subsequent executions. In our setup, this required $59.7$ hours for \tpcds\ and $26.5$ hours for \sqlstorm\ ($10$ executor configurations $\times\ 5$ repetitions per benchmark). In production settings where workload and cluster characteristics are stable, this cost is small relative to the cumulative cost savings during deployment. Where workloads drift gradually, models can be updated incrementally from newly observed executions, avoiding full retraining.

\paragraph{Integration Overhead.}
Our implementation extends the Mira resource manager~\cite{kaufmann2018mira} and requires only that the scheduler honor per-stage executor caps. The prediction and optimization steps run in milliseconds per stage and are queried at $300$\,ms intervals during execution, introducing negligible scheduling overhead.

%% file: sections/summary.tex
\section{Conclusion and Outlook}\label{sec:conclusion}

Efficient resource allocation remains a key challenge in distributed data processing systems, especially in serverless environments where cost and performance must be carefully balanced. This work presents a practical, data-driven approach to per-stage resource allocation in \Apache, offering substantial improvements over static, application-level strategies.
By training tree-based models to predict stage-level runtime and cost, and using these predictions to guide resource allocation, our framework enables user-defined \tradeoff s.
E.g., in \ONLINE\ mode with $\theta_r = 0.25$, we observe an average cost reduction of 58.6\% on \TPCDS\ at a 28.1\% runtime increase, compared to $\theta_r = 0.0$, which yields a 50.4\% cost reduction at a 16.4\% slowdown relative to the \Apache\ default.
This demonstrates that predictive modeling and user-defined trade-offs enable fine-grained, cost‑efficient optimization of large-scale data pipelines.

Our results are based on an empirical evaluation of \ScenarioC\ using a fixed set of \tpcds\ and \sqlstorm\ queries for both \zeroshot\ and \nshot\ approaches. Future work should broaden this evaluation to additional scenarios introduced in \Section~\ref{sec:solver} and investigate generalization to unseen queries.

\paragraph{Outlook on Generalization.}
In this work, the training data is limited to configurations with up to 128 executors (512 cores), which also defines the scope of our evaluation. 
We make no claims about model behavior outside the observed configuration range.
Changes in available resources or cluster configurations would require retraining the models to ensure reliable predictions and allocation decisions, as discussed in the previous section.
A promising direction for improving generalization beyond is to shift from stage-level to \emph{task}-level modeling, where predictions are based on more fine-grained execution characteristics that may extrapolate better across workloads and applications.
Incorporating task-level features and/or with hierarchical or multi-level learning frameworks could lead to more robust cost and performance estimates that generalize more effectively to unseen applications and workloads.

\paragraph{Outlook on Intelligent Scale-in.}
The performance metrics of this paper highlight the importance of intelligently allocating resources to a query's individual stages. Further cost reductions can be obtained from reducing resources during a query's execution. Perhaps simple ``timeout'' methods will be effective, however, the larger goal would be to preemptively scale-in resources based on estimated demand from subsequent stages. 

%% file: sections/appendix.tex
\newpage
\appendix

\section*{Appendix}

This appendix provides supplementary material that accompanies the paper, including extended experimental results, model details, and additional analyses.
Because of the workshop’s PDF submission size limit, all figures illustrating the additional results are provided in the \onlineappendix: \repolink.

\begin{itemize}
    \item \textit{Appendix~\ref{apx:code}: Code and Repository Details} --- Additional information on code release and additional artifacts.
    \item \textit{Appendix~\ref{apx:models}: Our Model Details} --- Additional information on the prediction models used in our evaluation.
    \item \textit{Appendix~\ref{apx:baselines}: Baselines} --- Additional information on the baseline methods used in our evaluation.
    \item \textit{Appendix~\ref{apx:results}: Extended Results} --- Extended analyses for TPC-DS and StormSQL benchmarks.
    \item \textit{Appendix~\ref{apx:hardware}: Hardware and Setup} --- Additional information on hardware and train/test setup.
\end{itemize}

\section{Code and Repository}\label{apx:code}

We are unable to release code due to corporate restrictions.
We do, however, provide the following artifacts in our \onlineappendix: \url{https://github.com/mrateike/vldb-submission}.
\begin{itemize}
    \item \textit{Results Data} --- Data used to generate the figures in the main paper and appendix (available in the \texttt{data} directory).
    \item \textit{Models} --- Trained prediction models used to generate results in the paper and appendix (available in the \texttt{models} directory).
\end{itemize}

%% file: sections/appendix-a-model-details.tex
\section{Our Prediction Models}\label{apx:models}

\subsection{Hyperparameters Best Models}

In Table~\ref{tab:ourshyperparams}, we report the hyperparameters for the GradientBoostingRegressor\footnote{\url{https://scikit-learn.org/stable/modules/generated/sklearn.ensemble.GradientBoostingRegressor.html}} models used in the main paper for end-to-end testing for Cost and Performance (\ExecutorRunTimeMax). The best hyperparameters have been selected using 5-fold-cross-validation. 

\begin{table*}[h]
\centering
\caption{Model hyperparameters and performance metrics for our best models used during evaluation.}
\label{tab:ourshyperparams}
\begin{tabular}{@{}p{1.2cm}p{1.2cm}p{7cm}cc@{}}
\toprule
\textbf{Data} & \textbf{Model} & \textbf{Hyperparameters} & \textbf{MSE} & \textbf{R²} \\
\midrule
\TPCDS & \ONLINE& \nestimators: 410, \maxdepthh: 6, \learningrate: 0.1, \subsample: 0.5, \minweightfractionleaf: 0.0 & $2.8957 \pm 0.1779$ & $0.5434 \pm 0.0156$ \\
\addlinespace
\TPCDS & \OFFLINE & \nestimators: 460, \maxdepthh: 7, \learningrate: 0.1, \subsample: 0.5, \minweightfractionleaf: 0.0& $0.5148 \pm 0.0385$ & $0.9186 \pm 0.0076$ \\
\addlinespace
\hline
\SQLStorm & \ONLINE & \nestimators: 460, \maxdepthh: 7, \learningrate: 0.1, \subsample: 0.5, \minweightfractionleaf: 0.0 & $2.6658 \pm 0.2016$ & $0.9400 \pm 0.0176$ \\
\addlinespace
\SQLStorm & \OFFLINE & \nestimators: 460, \maxdepthh: 6, \learningrate: 0.1, \subsample: 0.5, \minweightfractionleaf: 0.0 & $0.6524 \pm 0.5041$ & $0.9878 \pm 0.0061$ \\
\bottomrule
\end{tabular}
\end{table*}

\subsection{Cross-Validation of Our Models}

We report complete results for training \ONLINE\ and \OFFLINE\ models, comparing LinearRegression, GradientBoostingRegressor, and RandomForestRegressor for different targets, as explained in the main paper:

\begin{itemize}
    \item \Table~\ref{tab:apxtpcdsonline} shows \TPCDS\ \ONLINE\ model evaluation results
    \item \Table~\ref{tab:apxsqlstormonline} shows \SQLStorm\ \ONLINE\ model evaluation
    \item \Table~\ref{tab:apxtpcdsoffline} shows \TPCDS\ \OFFLINE\ model evaluation
    \item \Table~\ref{tab:apxsqlstormoffline} shows \SQLStorm\ \OFFLINE\ model evaluation
\end{itemize}

\begin{table*}[h]
\centering
\caption{\TPCDS\ \ONLINE\ model evaluation results. Linear Regression (\LinearRegression), Random Forest (\RandomForest), Gradient Boosting (\GradientBoosting). SD (DS), \ExecutorRunTimeMax (Max).}
\label{tab:apxtpcdsonline}
\begin{tabular}{@{}llp{6cm}cc@{}}
\toprule
\textbf{Target} & \textbf{Model} & \textbf{Parameters} & \textbf{MSE} & \textbf{R²} \\
\midrule
Cost & \LinearRegression & \fitintercept: False & $12{,}765{,}895.7817 \pm 5{,}101{,}713.6907$ & $-106.5049 \pm 43.5657$ \\
\addlinespace
Cost & \RandomForest & \bootstrap: True, \nestimators: 160, \maxdepthh: 7 & $45{,}759.9685 \pm 4{,}042.5299$ & $0.6174 \pm 0.0187$ \\
\addlinespace
Cost & \GradientBoosting & \nestimators: 210, \maxdepthh: 6, \learningrate: 0.1, \subsample: 0.5, \minweightfractionleaf: 0.0 & $38{,}652.0878 \pm 3{,}158.7891$ & $0.6766 \pm 0.0149$ \\
\addlinespace
\midrule
Perf (SD) & \LinearRegression & \fitintercept: False & $87.2132 \pm 22.3929$ & $-3.2026 \pm 1.1611$ \\
\addlinespace
Perf (SD) & \RandomForest & \bootstrap: True, \nestimators: 410, \maxdepthh: 7 & $11.7462 \pm 0.5864$ & $0.4377 \pm 0.0116$ \\
\addlinespace
Perf (SD) & \GradientBoosting & \nestimators: 410, \maxdepthh: 6, \learningrate: 0.1, \subsample: 0.5, \minweightfractionleaf: 0.0 & $10.2669 \pm 0.5193$ & $0.5085 \pm 0.0103$ \\
\addlinespace
\midrule
Perf (Max) & \LinearRegression & \fitintercept: True & $27.6213 \pm 5.3687$ & $-3.3617 \pm 0.8546$ \\
\addlinespace
Perf (Max) & \RandomForest & \bootstrap: True, \nestimators: 160, \maxdepthh: 7 & $3.0474 \pm 0.1948$ & $0.5195 \pm 0.0175$ \\
\addlinespace
Perf (Max) & \GradientBoosting & \nestimators: 410, \maxdepthh: 6, \learningrate: 0.1, \subsample: 0.5, \minweightfractionleaf: 0.0 & $2.8957 \pm 0.1779$ & $0.5434 \pm 0.0156$ \\
\bottomrule
\end{tabular}
\end{table*}

\begin{table*}
\caption{\SQLStorm\ \ONLINE\ model evaluation. Linear Regression (\LinearRegression), Random Forest (\RandomForest), Gradient Boosting (\GradientBoosting). SD (DS), \ExecutorRunTimeMax (Max).}
\label{tab:apxsqlstormonline}
\centering
\begin{tabular}{@{}llp{6cm}cc@{}}
\toprule
\textbf{Target} & \textbf{Model} & \textbf{Parameters} & \textbf{MSE} & \textbf{R²} \\
\midrule
Cost & \LinearRegression & \fitintercept: True & $924{,}778.1087 \pm 244{,}326.3015$ & $-2.9266 \pm 0.7885$ \\
\addlinespace
Cost & \RandomForest & \bootstrap: True, \nestimators: 10, \maxdepthh: 7 & $13{,}358.5137 \pm 846.0770$ & $0.9416 \pm 0.0115$ \\
\addlinespace
Cost & \GradientBoosting & \nestimators: 260, \maxdepthh: 7, \learningrate: 0.1, \subsample: 0.5, \minweightfractionleaf: 0.0 & $11{,}545.1384 \pm 358.1276$ & $0.9498 \pm 0.0075$ \\
\addlinespace
\midrule
Perf (DS) & \LinearRegression & \fitintercept: True & $58.4296 \pm 14.6535$ & $0.0816 \pm 0.0072$ \\
\addlinespace
Perf (DS) & \RandomForest & \bootstrap: True, \nestimators: 10, \maxdepthh: 7 & $8.5018 \pm 1.2000$ & $0.8609 \pm 0.0255$ \\
\addlinespace
Perf (DS) & \GradientBoosting & \nestimators: 360, \maxdepthh: 7, \learningrate: 0.1, \subsample: 0.5, \minweightfractionleaf: 0.0 & $6.0936 \pm 0.2921$ & $0.8971 \pm 0.0284$ \\
\addlinespace
\midrule
Perf (Max) & \LinearRegression & \fitintercept: False & $60.9472 \pm 11.1733$ & $-0.3060 \pm 0.1907$ \\
\addlinespace
Perf (Max) & \RandomForest & \bootstrap: False, \nestimators: 10, \maxdepthh: 7 & $20.0779 \pm 7.2291$ & $0.5891 \pm 0.0637$ \\
\addlinespace
Perf (Max) & \GradientBoosting & \nestimators: 460, \maxdepthh: 7, \learningrate: 0.1, \subsample: 0.5, \minweightfractionleaf: 0.0 & $2.6658 \pm 0.2016$ & $0.9400 \pm 0.0176$ \\
\bottomrule
\end{tabular}
\end{table*}

\begin{table*}
\caption{\TPCDS\ \OFFLINE\ model results. Linear Regression (\LinearRegression), Random Forest (\RandomForest), Gradient Boosting (\GradientBoosting). SD (DS), \ExecutorRunTimeMax (Max).}
\label{tab:apxtpcdsoffline}
\centering
\begin{tabular}{@{}llp{6cm}cc@{}}
\toprule
\textbf{Target} & \textbf{Model} & \textbf{Parameters} & \textbf{MSE} & \textbf{R²} \\
\midrule
Cost & \LinearRegression & \fitintercept: False & $56{,}421{,}446.1955 \pm 31{,}701{,}008.5355$ & $-475.6542 \pm 270.6458$ \\
\addlinespace
Cost & \RandomForest & \bootstrap: True, \nestimators: 110, \maxdepthh: 7 & $21{,}867.7576 \pm 1{,}541.9805$ & $0.8164 \pm 0.0151$ \\
\addlinespace
Cost & \GradientBoosting & \nestimators: 460, \maxdepthh: 7, \learningrate: 0.1, \subsample: 0.5, \minweightfractionleaf: 0.0 & $4{,}592.4841 \pm 334.0984$ & $0.9614 \pm 0.0035$ \\
\addlinespace
\midrule
Perf. (SD) & \LinearRegression & \fitintercept: False & $136.5464 \pm 33.6868$ & $-5.5600 \pm 1.6930$ \\
\addlinespace
Perf. (SD) & \RandomForest & \bootstrap: True, \nestimators: 110, \maxdepthh: 7 & $6.1407 \pm 0.3343$ & $0.7061 \pm 0.0079$ \\
\addlinespace
Perf. (SD) & \GradientBoosting & \nestimators: 460, \maxdepthh: 7, \learningrate: 0.1, \subsample: 0.5, \minweightfractionleaf: 0.0 & $1.9204 \pm 0.1576$ & $0.9079 \pm 0.0080$ \\
\addlinespace
\midrule
Perf. (Max) & \LinearRegression & \fitintercept: True & $174.3614 \pm 46.5823$ & $-26.7242 \pm 8.0328$ \\
\addlinespace
Perf. (Max) & \RandomForest & \bootstrap: True, \nestimators: 110, \maxdepthh: 7 & $1.2151 \pm 0.0902$ & $0.8077 \pm 0.0190$ \\
\addlinespace
Perf. (Max) & \GradientBoosting & \nestimators: 460, \maxdepthh: 7, \learningrate: 0.1, \subsample: 0.5, \minweightfractionleaf: 0.0 & $0.5148 \pm 0.0385$ & $0.9186 \pm 0.0076$ \\
\bottomrule
\end{tabular}
\end{table*}

\begin{table*}
\centering
\caption{\SQLStorm\ \OFFLINE\ model results. Linear Regression (\LinearRegression), Random Forest (\RandomForest), Gradient Boosting (\GradientBoosting). \StageDuration (SD), \ExecutorRunTimeMax (Max).}
\label{tab:apxsqlstormoffline}
\begin{tabular}{@{}llp{6cm}cc@{}}
\toprule
\textbf{Target} & \textbf{Model} & \textbf{Parameters} & \textbf{MSE} & \textbf{R²} \\
\midrule
Cost & \LinearRegression & \fitintercept: True & $1{,}105{,}307.8681 \pm 255{,}780.1731$ & $-3.7082 \pm 0.8108$ \\
\addlinespace
Cost & \RandomForest & \bootstrap: True, \nestimators: 460, \maxdepthh: 7 & $4{,}958.8561 \pm 363.9871$ & $0.9783 \pm 0.0042$ \\
\addlinespace
Cost & \GradientBoosting & \nestimators: 410, \maxdepthh: 7, \learningrate: 0.1, \subsample: 0.5, \minweightfractionleaf: 0.0 & $856.2648 \pm 522.6476$ & $0.9962 \pm 0.0025$ \\
\addlinespace
\midrule
Perf. (SD) & \LinearRegression & \fitintercept: True & $56.7856 \pm 13.1241$ & $0.1033 \pm 0.0300$ \\
\addlinespace
Perf. (SD) & \RandomForest & \bootstrap: True, \nestimators: 110, \maxdepthh: 7 & $6.0071 \pm 0.7221$ & $0.9017 \pm 0.0169$ \\
\addlinespace
Perf. (SD) & \GradientBoosting & \nestimators: 460, \maxdepthh: 7, \learningrate: 0.1, \subsample: 0.5, \minweightfractionleaf: 0.0 & $2.2269 \pm 0.3552$ & $0.9636 \pm 0.0064$ \\
\addlinespace
Perf. (Max) & \LinearRegression & \fitintercept: True & $65.6093 \pm 10.3240$ & $-0.4159 \pm 0.2239$ \\
\addlinespace
\midrule
Perf. (Max) & \RandomForest & \bootstrap: True, \nestimators: 10, \maxdepthh: 7 & $8.5395 \pm 3.5709$ & $0.8305 \pm 0.0439$ \\
\addlinespace
Perf. (Max) & \GradientBoosting & \nestimators: 460, \maxdepthh: 6, \learningrate: 0.1, \subsample: 0.5, \minweightfractionleaf: 0.0 & $0.6524 \pm 0.5041$ & $0.9878 \pm 0.0061$ \\
\bottomrule
\end{tabular}
\end{table*}

%% file: sections/appendix-b-baselines.tex
\section{Baseline Models}\label{apx:baselines}

For all cross-validation experiments for our models and baselines we used the same training/testing splits.

\subsection{Adaptations}

\paragraph{ReLoca Adaptation (\ReLocaAdapt)}

\begin{itemize}
    \item \textit{Category:} n-shot
    \item \textit{Key idea:} ReLoca is a deep learning-based framework that optimizes resource allocation for data-parallel jobs to minimize job completion time. It employs a deep neural network trained with an adaptive sampling method to learn the impact of job operations on system overhead and compute time, achieving a 29.85\% reduction in job completion time compared to existing methods.
    \item \textit{Original setting:} ReLoca focuses in optimizing performance rather than cost at an application level. The computational domains differ substantially between the two approaches---ReLoca targets machine learning workloads, whereas our evaluation utilizes SQL query execution traces.
    \item \textit{Adaptations considerations:} We adopted the input features specified in Table 1 from the original ReLoca publication and incorporated the proposed neural network architecture (as depicted in Figure 5) into our hyperparameter optimization framework. Following the computational methodology described in Figure 4, we derived the DAG width and depth metrics for each application. A fundamental adaptation was required due to our framework's stage-level granularity for resource recommendation: we transitioned from application-level execution duration prediction (as originally formulated in ReLoca) to stage-level predictions.
    
    Regarding data collection and experimental design, ReLoca employs an interventional sampling protocol that systematically balances under-allocation scenarios (which degrade performance) and over-allocation scenarios (which introduce substantial system overhead). This approach yields approximately 600 samples per application (approximately 3,000 total experimental runs), with 200 samples per application reserved for training.
    
    In contrast, our experimental dataset comprises approximately 50 runs per benchmark (totaling 100 experimental runs across both benchmarks), substantially constrained by data collection limitations inherent to production system monitoring. These methodological disparities in data scale, workload characteristics, prediction granularity, and experimental scope, represent significant constraints in our experimental setup and may impact the direct comparability of baseline method performance in our evaluation.
    \item \textit{Grid search parameters:} The final model configuration was selected through a grid search procedure that systematically evaluated the hyperparameter space across five dimensions: hidden layer sizes and depth, activation functions, learning rate parameters, maximum iteration thresholds, and solver algorithms.
\end{itemize}

\paragraph{CherryPick Adaptation (\CherryPickAdapt)}

\begin{itemize}
    \item \textit{Category:} n-shot
    \item \textit{Key idea:} CherryPick addresses the challenge of selecting optimal cloud configurations for recurring big data analytics jobs. The system employs Bayesian Optimization to construct performance models that achieve sufficient accuracy to identify the best or near-best configurations from numerous VM instance types and cluster sizes with minimal test runs. By building models that distinguish optimal configurations from suboptimal alternatives through only a few experimental evaluations, CherryPick significantly reduces the search cost overhead. Experimental validation demonstrates that CherryPick achieves a 45-90\% success rate in identifying optimal configurations, while reducing search cost by up to 75\% compared to existing configuration selection approaches.
    \item \textit{Original setting:} CherryPick is a solution to predict multiple resources at VM and application level, this include recommending CPU, Disk, RAM, Network resources, their main objective is to reduce cost. This means their recommendations are static to create the VM, not a dynamic recommendation of resources like is proposed in our paper. They include SQL applications, such as \TPCDS and TPC-H, as well as machine learning workloads.
    \item \textit{Adaptations considerations:} We adopted the Gaussian Process formulation employed by CherryPick in their Bayesian Optimization engine (Section 3.5 in the original publication) as the foundational approach for cost estimation. However, we adapted the input feature space to reflect our system's architectural constraints: rather than utilizing global VM configuration parameters (instance type, CPU allocation, network bandwidth, memory allocation), we employed stage-level performance indicators including task count, executor allocation, partition sizes, and related metrics to estimate cost (in our case the target variable \texttt{ExecutorRunTime}). This transition represents a substantial distribution shift in the feature space and data characteristics compared to CherryPick's formulation, while maintaining the same modeling approach. CherryPick's training dataset comprises interventional data derived from 66 distinct cloud configurations, systematically varying VM instance types, CPU cores, network specifications, and memory allocations (as detailed in Section 5.1 from their paper). In contrast, our evaluation utilizes 10 configuration variants, reflecting constraints in experimental data collection feasibility. Furthermore, the resource recommendation mechanisms differ fundamentally: CherryPick employs a Bayesian Optimization allocation process that leverages the Gaussian Process prior for configuration recommendation, whereas our approach implements the cost-effective solver described in Section 3.4 of our paper with $\theta_r = 0.0$. These methodological differences in data variations, workload characteristics, prediction granularity, configuration space, and resource allocation methodology represent significant constraints in our experimental setup and may impact the performance of our adaptation relative to CherryPick approach.
    \item \textit{Grid search parameters:} The final model configuration was selected through a grid search procedure that systematically evaluated the hyperparameter space across four critical dimensions: regularization parameter (alpha), target variable normalization strategy, optimizer restart behavior, and kernel specifications.
\end{itemize}

\subsection{Input Features}

\textit{\tpcds\ \ReLocaAdapt\ features}: Number of Tasks, NumberExecutors, InputBytesperStage, PartitionsperStage, NumberOperations, DAGwidth, DAGdepth, DataSourceRDD, FileScanRDD, MapPartitions- \linebreak RDD, ParallelCollectionRDD, ShuffledRowRDD, UnionRDD, ZippedPartitionsRDD2, CartesianColumnarBatchRDD, GlutenWholeStage- ColumnarRDD, ShuffledColumnarBatchRDD, WholeStageZippedPartitionsRDD.

\textit{\sqlstorm\ \ReLocaAdapt\ features}: Number of Tasks, ExecutorCores, NumberExecutors, InputBytesperStage, PartitionsperStage, NumberOperations, DAGwidth, DAGdepth, DataSourceRDD, FileScanRDD, MapPartitionsRDD, ParallelCollectionRDD, ShuffledRowRDD, UnionRDD, ZippedPartitionsRDD2, CartesianColumnarBatchRDD, GlutenWholeStageColumnarRDD, ShuffledColumnarBatchRDD, \linebreak WholeStageZippedPartitionsRDD.

\textit{\CherryPickAdapt\ features, both \tpcds\ and \sqlstorm}: Number of Tasks, NumberExecutors, InputBytesperStage, PartitionsperStage, NumberOperations, DataSourceRDD, FileScanRDD, MapPartitionsRDD, ParallelCollectionRDD, ShuffledRowRDD, UnionRDD, \linebreak ZippedPartitionsRDD2, CartesianColumnarBatchRDD, Gluten- \linebreak WholeStageColumnarRDD, ShuffledColumnarBatchRDD, Whole- \linebreak StageZippedPartitionsRDD.

\begin{table*}[]
\caption{Baseline hyperparameter configurations and performance metrics. DurationStage (DS), ExecutorRunTimeMaxbyCore (Max). \CherryPickAdapt\ (CP) and \ReLocaAdapt\ (RLo).}
    \centering
\begin{tabular}{@{}p{1.2cm}p{0.7cm}p{0.8cm}p{0.8cm}p{4.3cm}p{1.1cm}p{1.1cm}p{1.1cm}p{1.1cm}p{1.1cm}p{1.1cm}@{}}
\toprule
\textit{Dataset} & \textit{Model} & \textit{Target} & \textit{Log} & \textit{Parameters} & \textit{Train} & \textit{Test} & \textit{Mean MSE} & \textit{Mean R$^2$} & \textit{Best MSE} & \textit{Best R$^2$} \\
\midrule
\tpcds & CP & DS & True & alpha: 1e-06, normalize\_y: true, n\_restarts\_optimizer: 1 &  29,721 & 7,431 & 8.04 & 0.829 & 7.25 & 0.834 \\
\addlinespace
\tpcds & RLo & Max & False & hidden\_layer\_sizes: [20,10,10,10,10], activation: logistic, learning\_rate\_init: 0.0001, max\_iter: 1000, solver: adam &  49,580 & 12,395 & 27.11 & 0.335 & 13.42 & 0.591 \\
\addlinespace
\tpcds & CP & Max & True & alpha: 1e-06, normalize\_y: true, n\_restarts\_optimizer: 1 &  29,721 & 7,431 & 3.34 & 0.893 & 2.81 & 0.938 \\
\addlinespace
\tpcds & RLo & DS & False & hidden\_layer\_sizes: [20,10,10,10,10], activation: tanh, learning\_rate\_init: 0.0001, max\_iter: 1000, solver: adam &  49,580 & 12,395 & 44.22 & 0.207 & 36.07 & 0.242 \\
\addlinespace
\sqlstorm & CP & Max & True & alpha: 1e-06, normalize\_y: true, n\_restarts\_optimizer: 1 &  49,580 & 12,395 & 6.55 & 0.798 & 3.35 & 0.932 \\
\addlinespace
\sqlstorm & RLo & DS & False & hidden\_layer\_sizes: [10,5,5,5,5], activation: relu, learning\_rate\_init: 0.001, max\_iter: 200, solver: adam &  19,858 & 4,965 & 124.52 & -1.233 & 56.14 & -0.001 \\
\addlinespace
\sqlstorm & CP & DS & True & alpha: 1e-06, normalize\_y: true, n\_restarts\_optimizer: 1 &  49,580 & 12,395 & 18.16 & 0.662 & 10.55 & 0.776 \\
\addlinespace
\sqlstorm & RLo & Max & False & hidden\_layer\_sizes: [10,5,5,5,5], activation: relu, learning\_rate\_init: 0.001, max\_iter: 200, solver: adam &  19,858 & 4,965 & 103.21 & -1.540 & 41.15 & 0.003 \\
\bottomrule
\end{tabular}
\end{table*}

\subsection{Model Availability}

The trained regression models and their configurations are available in our reposity\footnote{\repolink} \texttt{models/models\_baselines} directory, organized by dataset (\tpcds\ and \sqlstorm). Each model is stored as a ONNX and JSON files. The JSON files contain comprehensive model metadata including: the model name and type, hyperparameters specific to each model architecture, cross-validation performance metrics, information about negative predictions, the best model instance metrics, target column name, run type classification, log transformation flag, feature column names used for training, training and test dataset sizes, and folder organization information. This structured format enables easy retrieval and deployment of models while maintaining full reproducibility of the modeling configurations. The feature sets include computational characteristics such as task counts, executor configuration, data volumes, partition information, operation counts, DAG topology metrics, and RDD type distributions, providing comprehensive coverage of the factors influencing query performance and resource utilization. The ONNX file contains the trained model ready to use at inference time.

\subsection{Training Time Considerations}

While this work compares performance and cost across methods, we need to note that training time can differ substantially between the implemented methods. On a M1 Max, 64GB RAM laptop, \ReLocaAdapt~\cite{hu2020reloca} trained in $\approx 12$ seconds on both benchmarks, similar to our approaches, which require $\approx 11$ seconds for both benchmarks. In contrast, \CherryPickAdapt~\cite{alipourfard2017cherrypick} required $\approx 3$ hours $39$ minutes on \sqlstorm\ and $\approx 8$ hours $54$ minutes on \TPCDS.

%% file: sections/appendix-c-extended-results.tex
\section{Extended Results}\label{apx:results}

Due to the aforementioned workshop’s PDF submission size limit, we provide (i) additional SHAP~\cite{SHAP_NeurIPS} plots to analyze feature importance and influence for both our \emph{cost} and \emph{performance} prediction models across benchmarks and training regimes, and (ii) end‑to‑end testing Scenario $\gamma$ using models that predict \DurationStage\ in the \onlineappendix: \repolink.

\paragraph{Extended Per Query Results}

We report additional results on the per query analyses in both benchmarks, namely the top-5 queries with the largest divergence in cost from $0$-shot and \Apache\ approach for both benchmarks:

\begin{itemize}
    \item Table~\ref{tab:tpcds-top5}: top-5 queries for \tpcds
    \item Table~\ref{tab:sqlstorm-top5}: top-5 queries for \sqlstorm
\end{itemize}

Additional radar plots for cost and performance across all queries and benchmarks can be found in the \onlineappendix.

\begin{table}[h]
\centering
\caption{Top-5 queries with largest cost divergence, TPC-DS}
\label{tab:tpcds-top5}
\begin{tabular}{@{}lrrr@{}}
\toprule
\textbf{Query} & \textbf{Cost 0-shot} & \textbf{Cost \Apache} & \textbf{Cost Difference} \\
\midrule
14b & 7,209,175 & 16,333,383 & 9,124,208 \\
03 & 528,223 & 9,253,531 & 8,725,308 \\
14a & 6,705,714 & 14,700,672 & 7,994,958 \\
23b & 5,357,945 & 9,911,296 & 4,553,351 \\
04 & 6,476,707 & 10,448,988 & 3,972,281 \\
\bottomrule
\end{tabular}
\end{table}

\begin{table}[h]
\centering
\caption{Top-5 queries w. largest cost divergence, SQLStorm}
\label{tab:sqlstorm-top5}
\begin{tabular}{@{}lrrr@{}}
\toprule
\textbf{Query} & \textbf{Cost 0-shot} & \textbf{Cost \Apache} & \textbf{Cost Difference} \\
\midrule
3053 & 1,416,380 & 5,810,624 & 4,394,244 \\
1296 & 3,088,394 & 5,723,702 & 2,635,308 \\
3020 & 910,406 & 3,111,072 & 2,200,666 \\
23239 & 694,750 & 2,267,374 & 1,572,624 \\
749 & 561,207 & 1,674,908 & 1,113,701 \\
\bottomrule
\end{tabular}
\end{table}

\paragraph{Data Resources}
In the \onlineappendix\footnote{\repolink},  \texttt{data/perf-cost\_} \texttt{tcpds\_baselines\_duration.csv} and \texttt{data/perf-cost\_tcpds\_} \texttt{baselines\_maxbycore.csv} contain aggregated statistical summar-ies of performance and cost metrics for the \tpcds\ benchmark comparing different strategies. 
Columns provide mean and standard deviation values for both Performance (Perf. mean, Perf. std) and Cost (Cost mean, Cost std) metrics across multiple query executions. 
The dataset includes six optimization strategies: \zeroshot\ approaches with varying theta parameters ($\theta_r=0.0$, 0.25, and 0.5), three \nshot\ approaches with corresponding theta settings ($\theta_r=0.0$, 0.25, and 0.5), and the \Apache\ baseline for comparison.

\texttt{data/perf-cost\_sqlstorm\_maxbycore\_per\_}\texttt{query.csv} con- \linebreak tains performance and cost metrics for the \sqlstorm\ benchmark across multiple proposed methods and baseline strategies. Each record includes four key metrics: Performance (Perf.), Cost, both in milliseconds, Query identifier, and method (optimization strategy). The dataset encompasses multiple run methods including 0-shot/THETA-0.0, 0-shot/THETA-0.25, 0-shot/THETA-0.5, n-shot/ \linebreak THETA-0.0, n-shot/THETA-0.25, n-shot/THETA-0.5, VANILLA baseline, and CHERRY-PICK strategies. Where 0-shot = $0$-shot approach, n-shot = $n$-shot approach, VANILLA = \Apache\ run. The data represents multiple executions per query (indicated by repeated Query IDs with varying Perf. and Cost values), allowing for analysis of performance variability and consistency across different approaches. Similarly \texttt{data/perf-cost\_tpcds\_maxbycore\_per\_} \texttt{query.csv} contains same metrics for \tpcds.

%% file: sections/appendix-d-hardware-setup.tex
\section{Hardware and Train/Test Setup}\label{apx:hardware}

\paragraph{Cluster Hardware Configuration}

The data collection and model evaluation hardware configuration includes 20 OpenShift worker nodes, each equipped with 4 x Intel Xeon E5-2683 v4 CPUs running at 2.10 GHz (16 cores per CPU, 64 physical cores per node), 1.5 TB of RAM, and 128 GB NVMe local storage. Hyper-Threading is enabled, providing 128 logical CPUs per node. The nodes are interconnected using dual 25 Gbit Ethernet links. Persistent data is stored on an S3-compatible object store accessed via the S3A connector.

\paragraph{Data Processing Infrastructure}

Our data processing infrastructure leverages \Apache\ 3.5.4 integrated with Iceberg as the underlying table format. Iceberg is configured with a Hive-backed catalog and an external Hive Metastore for metadata management while table data is stored in S3-compatible object storage accessed via the S3A connector. The execution engine is accelerated using Gluten with the Velox native backend, enabling columnar execution and native vectorized processing.

\paragraph{Train/Test Configurations}

For Training, we shut down an executor after 60 seconds of idleness, following the default \Apache\ configuration. The maximum available number of executors is 128. For testing, \Apache\ queries Mira for updates on the recommended number of cores computed by our optimization framework at 300-millisecond intervals. Multiple stages may run in parallel as long as their combined recommended executors remain within a limit of 128 executors. For our methods and \CherryPickAdapt, any executor that becomes idle is shut down immediately when no further tasks remain for its active stage. The \Apache\ baseline uses the default 60-second executor idle timeout.

%% file: references.bib
@inproceedings{kaufmann2018mira,
  title={Mira: sharing resources for distributed analytics at small timescales},
  author={Kaufmann, Michael and Kourtis, Kornilios and Schuepbach, Adrian and Zitterbart, Martina},
  booktitle={2018 IEEE International Conference on Big Data (Big Data)},
  pages={231--241},
  year={2018},
  organization={IEEE}
}

@article{lyu2022fine,
  title={Fine-grained modeling and optimization for intelligent resource management in big data processing},
  author={Lyu, Chenghao and Fan, Qi and Song, Fei and Sinha, Arnab and Diao, Yanlei and Chen, Wei and Ma, Li and Feng, Yihui and Li, Yaliang and Zeng, Kai and others},
  journal={Proceedings of the VLDB Endowment},
  volume={15},
  number={11},
  pages={3098--3111},
  year={2022},
  publisher={VLDB Endowment}
}

@inproceedings{dimopoulos2017justice,
  title={Justice: A deadline-aware, fair-share resource allocator for implementing multi-analytics},
  author={Dimopoulos, Stratos and Krintz, Chandra and Wolski, Rich},
  booktitle={2017 IEEE International Conference on Cluster Computing (CLUSTER)},
  pages={233--244},
  year={2017},
  organization={IEEE}
}

@inproceedings{venkataraman2016ernest,
  title={Ernest: Efficient performance prediction for $\{$Large-Scale$\}$ advanced analytics},
  author={Venkataraman, Shivaram and Yang, Zongheng and Franklin, Michael and Recht, Benjamin and Stoica, Ion},
  booktitle={13th USENIX symposium on networked systems design and implementation (NSDI 16)},
  pages={363--378},
  year={2016}
}

@inproceedings{sidhanta2016optex,
  title={Optex: A deadline-aware cost optimization model for spark},
  author={Sidhanta, Subhajit and Golab, Wojciech and Mukhopadhyay, Supratik},
  booktitle={2016 16th IEEE/ACM International Symposium on Cluster, Cloud and Grid Computing (CCGrid)},
  pages={193--202},
  year={2016},
  organization={IEEE}
}

@inproceedings{wang2016novel,
  title={A novel method for tuning configuration parameters of spark based on machine learning},
  author={Wang, Guolu and Xu, Jungang and He, Ben},
  booktitle={2016 IEEE 18th International Conference on High Performance Computing and Communications; IEEE 14th International Conference on Smart City; IEEE 2nd International Conference on Data Science and Systems (HPCC/SmartCity/DSS)},
  pages={586--593},
  year={2016},
  organization={IEEE}
}

@inproceedings{siddiqui2020cost,
  title={Cost models for big data query processing: Learning, retrofitting, and our findings},
  author={Siddiqui, Tarique and Jindal, Alekh and Qiao, Shi and Patel, Hiren and Le, Wangchao},
  booktitle={Proceedings of the 2020 ACM SIGMOD International Conference on Management of Data},
  pages={99--113},
  year={2020}
}

@inproceedings{
casacuberta2026good,
title={Good Allocations from Bad Estimates},
author={S{\'\i}lvia Casacuberta and Moritz Hardt},
booktitle={The Fourteenth International Conference on Learning Representations},
year={2026},
url={https://openreview.net/forum?id=rxZdaKhu2I}
}

@article{SHAP_NeurIPS,
  title={A Unified Approach to Interpreting Model Predictions},
  author={Lundberg, Scott M and Lee, Su-In},
  journal={Advances in Neural Information Processing Systems},
  volume={30},
  year={2017}
}

@inproceedings{liu2025supporting,
  title={Supporting our ai overlords: Redesigning data systems to be agent-first},
  author={Liu, Shu and Ponnapalli, Soujanya and Shankar, Shreya and Zeighami, Sepanta and Zhu, Alan and Agarwal, Shubham and Chen, Ruiqi and Suwito, Samion and Yuan, Shuo and Stoica, Ion and others},
booktitle={6th Annual Conference on
Innovative Data Systems Research (CIDR ’26)},
  year={2026}
}

@article{lundberg2020local2global,
  title={From local explanations to global understanding with explainable AI for trees},
  author={Lundberg, Scott M. and Erion, Gabriel and Chen, Hugh and DeGrave, Alex and Prutkin, Jordan M. and Nair, Bala and Katz, Ronit and Himmelfarb, Jonathan and Bansal, Nisha and Lee, Su-In},
  journal={Nature Machine Intelligence},
  volume={2},
  number={1},
  pages={2522-5839},
  year={2020},
  publisher={Nature Publishing Group}
}

@article{sqlstorm25,
  author  = {Tobias Schmidt and Viktor Leis and Peter Boncz and Thomas Neumann},
  title   = {SQLStorm: Taking Database Benchmarking into the LLM Era},
  journal = {Proceedings of the VLDB Endowment},
  volume  = {18},
  number  = {11},
  pages   = {4144--4157},
  year    = {2025},
  publisher={VLDB Endowment}
}

@inproceedings{sewal2022machine,
  title={A machine learning approach for predicting execution statistics of spark application},
  author={Sewal, Piyush and Singh, Hari},
  booktitle={2022 Seventh International Conference on Parallel, Distributed and Grid Computing (PDGC)},
  pages={331--336},
  year={2022},
  organization={IEEE}
}

@inproceedings{maros2019machine,
  title={Machine learning for performance prediction of spark cloud applications},
  author={Maros, Alexandre and Murai, Fabricio and da Silva, Ana Paula Couto and Almeida, Jussara M and Lattuada, Marco and Gianniti, Eugenio and Hosseini, Marjan and Ardagna, Danilo},
  booktitle={2019 IEEE 12th International Conference on Cloud Computing (CLOUD)},
  pages={99--106},
  year={2019},
  organization={IEEE}
}

@inproceedings{hu2020reloca,
  title={Reloca: Optimize resource allocation for data-parallel jobs using deep learning},
  author={Hu, Zhiyao and Li, Dongsheng and Zhang, Dongxiang and Chen, Yixin},
  booktitle={IEEE INFOCOM 2020-IEEE Conference on Computer Communications},
  pages={1163--1171},
  year={2020},
  organization={IEEE}
}

@article{hu2021optimizing,
  title={Optimizing resource allocation for data-parallel jobs via GCN-based prediction},
  author={Hu, Zhiyao and Li, Dongsheng and Zhang, Dongxiang and Zhang, Yiming and Peng, Baoyun},
  journal={IEEE Transactions on Parallel and Distributed Systems},
  volume={32},
  number={9},
  pages={2188--2201},
  year={2021},
  publisher={IEEE}
}

@inproceedings{thanopoulou2012benchmarking,
  title={Benchmarking with TPC-H on off-the-shelf hardware an experiments report},
  author={Thanopoulou, Anna and Carreira, Paulo and Galhardas, Helena},
  booktitle={14th International Conference on Enterprise Information Systems, ICEIS 2012},
  pages={205--208},
  year={2012}
}

@article{barata2015overview,
  title={An overview of decision support benchmarks: TPC-DS, TPC-H and SSB},
  author={Barata, Melyssa and Bernardino, Jorge and Furtado, Pedro},
  journal={New Contributions in Information Systems and Technologies: Volume 1},
  pages={619--628},
  year={2015},
  publisher={Springer}
}

@inproceedings{nambiar2006making,
  title={The making of tpc-ds},
  author={Nambiar, Raghunath Othayoth and Poess, Meikel},
  booktitle={Proceedings of the International Conference on Very Large Data Bases},
  volume={32},
  number={2},
  pages={1049--1058},
  year={2006}
}

@inproceedings{poess2007you,
  title={Why you should run TPC-DS: a workload analysis},
  author={Poess, Meikel and Nambiar, Raghunath Othayoth and Walrath, David},
  booktitle={Proceedings of the 33rd international conference on Very large data bases},
  pages={1138--1149},
  year={2007}
}

@inproceedings{tariq2023execution,
  title={Execution time prediction model that considers dynamic allocation of spark executors},
  author={Tariq, Hina and Das, Olivia},
  booktitle={European Workshop on Performance Engineering},
  pages={340--352},
  year={2023},
  organization={Springer}
}

@inproceedings{alipourfard2017cherrypick,
  title={$\{$CherryPick$\}$: Adaptively unearthing the best cloud configurations for big data analytics},
  author={Alipourfard, Omid and Liu, Hongqiang Harry and Chen, Jianshu and Venkataraman, Shivaram and Yu, Minlan and Zhang, Ming},
  booktitle={14th USENIX Symposium on Networked Systems Design and Implementation (NSDI 17)},
  pages={469--482},
  year={2017}
}

@article{ahmed2021enhanced,
  title={An enhanced parallelisation model for performance prediction of apache spark on a multinode hadoop cluster},
  author={Ahmed, Nasim and Barczak, Andre LC and Rashid, Mohammad A and Susnjak, Teo},
  journal={Big Data and Cognitive Computing},
  volume={5},
  number={4},
  pages={65},
  year={2021},
  publisher={MDPI}
}

@article{zaharia2016apache,
  title={Apache spark: a unified engine for big data processing},
  author={Zaharia, Matei and Xin, Reynold S and Wendell, Patrick and Das, Tathagata and Armbrust, Michael and Dave, Ankur and Meng, Xiangrui and Rosen, Josh and Venkataraman, Shivaram and Franklin, Michael J and others},
  journal={Communications of the ACM},
  volume={59},
  number={11},
  pages={56--65},
  year={2016},
  publisher={ACM New York, NY, USA}
}

@article{scikit-learn,
  title={Scikit-learn: Machine Learning in {P}ython},
  author={Pedregosa, F. and Varoquaux, G. and Gramfort, A. and Michel, V.
          and Thirion, B. and Grisel, O. and Blondel, M. and Prettenhofer, P.
          and Weiss, R. and Dubourg, V. and Vanderplas, J. and Passos, A. and
          Cournapeau, D. and Brucher, M. and Perrot, M. and Duchesnay, E.},
  journal={Journal of Machine Learning Research},
  volume={12},
  pages={2825--2830},
  year={2011}
}

@inproceedings{gao2017autopath,
  title={AutoPath: harnessing parallel execution paths for efficient resource allocation in multi-stage big data frameworks},
  author={Gao, Han and Yang, Zhengyu and Bhimani, Janki and Wang, Teng and Wang, Jiayin and Sheng, Bo and Mi, Ningfang},
  booktitle={2017 26th International Conference on Computer Communication and Networks (ICCCN)},
  pages={1--9},
  year={2017},
  organization={IEEE}
}

@inproceedings{brewer2015kubernetes,
author = {Brewer, Eric A.},
title = {Kubernetes and the path to cloud native},
year = {2015},
isbn = {9781450336512},
publisher = {Association for Computing Machinery},
address = {New York, NY, USA},
url = {https://doi.org/10.1145/2806777.2809955},
doi = {10.1145/2806777.2809955},
booktitle = {Proceedings of the Sixth ACM Symposium on Cloud Computing},
pages = {167},
numpages = {1},
location = {Kohala Coast, Hawaii},
series = {SoCC '15}
}

@inproceedings{vavilapalli2013apache,
  title={Apache hadoop yarn: Yet another resource negotiator},
  author={Vavilapalli, Vinod Kumar and Murthy, Arun C and Douglas, Chris and Agarwal, Sharad and Konar, Mahadev and Evans, Robert and Graves, Thomas and Lowe, Jason and Shah, Hitesh and Seth, Siddharth and others},
  booktitle={Proceedings of the 4th Annual Symposium on Cloud Computing},
  pages={1--16},
  year={2013}
}

@article{gartner_cloud_budgets,
  title = {Why Cloud Budgets Don’t Stay in Check — and How to Make Sure Yours Do},
  author = {{Gartner, Inc.}},
  year = {2023},
  url = {https://www.gartner.com/en/articles/why-cloud-budgets-don-t-stay-in-check-and-how-to-make-sure-yours-do},
  note = {Accessed: 2025-06-27},
  journal = {Gartner Research}
}
